\documentclass[acmsmall]{acmart}
\AtBeginDocument{%
  \providecommand\BibTeX{{%
    \normalfont B\kern-0.5em{\scshape i\kern-0.25em b}\kern-0.8em\TeX}}}

\setcopyright{acmcopyright}
\copyrightyear{2018}
\acmYear{2018}
\acmDOI{XXXXXXX.XXXXXXX}

\acmJournal{JACM}
\acmVolume{37}
\acmNumber{4}
\acmArticle{111}
\acmMonth{8}

\usepackage{enumitem}
\usepackage{mathtools}
\usepackage{array}
\usepackage{multirow}
\usepackage{xcolor,soul}

\newcolumntype{H}{>{\setbox0=\hbox\bgroup}c<{\egroup}@{}}

\definecolor{lightgray}{RGB}{225, 225, 225}

\newcommand{\hlbox}[2]{
  \begin{center}
    \fcolorbox{white}{lightgray}{
      \parbox{.9\columnwidth}{\noindent \textbf{#1}. \textit{#2}}
    }
  \end{center}
}

\newcolumntype{P}[1]{>{\centering\arraybackslash}p{#1}}

\newcommand{\textBF}[1]{%
    \pdfliteral direct {2 Tr 0.3 w} %the second factor is the boldness
     #1%
    \pdfliteral direct {0 Tr 0 w}%
}

\theoremstyle{plain}
\newtheorem{definition}{Definition}
\newtheorem{problem}{Problem}

\begin{document}

% \title[Counterfactual Explanations of Unfairness in Recommendation]{Counterfactual Explanations of Unfairness for Graph Learning Representation in Recommender Systems}

\title[Unfairness Explanation in GNNs for Recommendation]{\color{black} GNNUERS: Fairness Explanation in GNNs for Recommendation via Counterfactual Reasoning}
% GNNUERS: Unfairness Explanation in Recommender Systems through Counterfactually-Perturbed Graphs}

\author{Giacomo Medda}
\orcid{0000-0002-1300-1876}
\affiliation{%
  \institution{University of Cagliari}
  \streetaddress{Via Ospedale, 72}
  \city{Cagliari}
  \country{Italy}
}
\email{giacomo.medda@unica.it}

\author{Francesco Fabbri}
\orcid{0000-0002-9631-1799}
\affiliation{%
  \institution{Spotify}
  \city{Barcelona}
  \country{Spain}
}
\email{francescof@spotify.com}

\author{Mirko Marras}
\orcid{0000-0003-1989-6057}
\affiliation{%
  \institution{University of Cagliari}
  \streetaddress{Via Ospedale, 72}
  \city{Cagliari}
  \country{Italy}
}
\email{mirko.marras@acm.org}

\author{Ludovico Boratto}
\orcid{0000-0002-6053-3015}
\affiliation{%
  \institution{University of Cagliari}
  \streetaddress{Via Ospedale, 72}
  \city{Cagliari}
  \country{Italy}
}
\email{ludovico.boratto@acm.org}

% \author{Mihnea Tufis}
% \orcid{0000-0003-4888-1639}
% \affiliation{%
%   \institution{Eurecat}
%   \streetaddress{}
%   \city{Barcelona}
%   \country{Spain}
% }
% \email{mihnea.tufis@eurecat.org}

\author{Gianni Fenu}
\orcid{0000-0003-1989-6057}
\affiliation{%
  \institution{University of Cagliari}
  \streetaddress{Via Ospedale, 72}
  \city{Cagliari}
  \country{Italy}
}
\email{fenu@unica.it}

\renewcommand{\shortauthors}{Medda et al.}

\begin{abstract}
  Nowadays, research into personalization has been focusing on explainability and fairness. Several approaches proposed in recent works are able to explain individual recommendations in a post-hoc manner or by explanation paths. However, explainability techniques applied to unfairness in recommendation have been limited to finding user/item features mostly related to biased recommendations. In this paper, we devised a novel algorithm that leverages counterfactuality methods to discover user unfairness explanations in the form of user-item interactions. In our counterfactual framework, interactions are represented as edges in a bipartite graph, with users and items as nodes. Our bipartite graph explainer perturbs the topological structure to find an altered version that minimizes the disparity in utility between the protected and unprotected demographic groups. Experiments on four real-world graphs coming from various domains showed that our method can systematically explain user unfairness on three state-of-the-art GNN-based recommendation models. Moreover, an empirical evaluation of the perturbed network uncovered relevant patterns that justify the nature of the unfairness discovered by the generated explanations. The source code and the preprocessed data sets are available at \url{https://github.com/jackmedda/RS-BGExplainer}.
\end{abstract}




\begin{teaserfigure}
    \centering
    \includegraphics[width=1.0\textwidth]{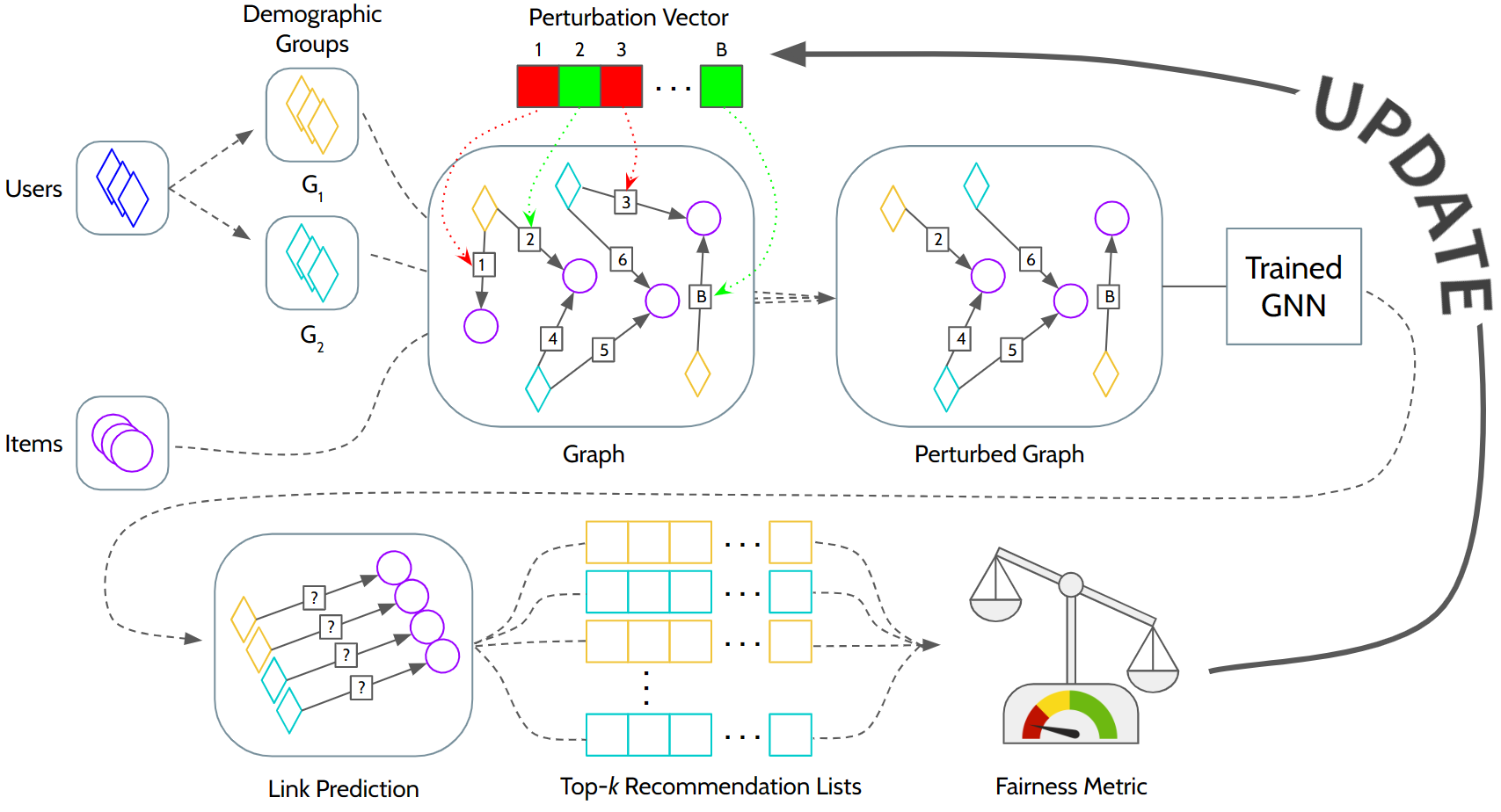}
    \caption{\texttt{GNNUERS} updates the perturbation vector such that the removed user-item interactions from the graph lead the trained GNN to generate fairer recommendations. The perturbation vector represents the counterfactual explanation of the prior unfairness across demographic groups.}
    \Description{figure description}
    \label{fig:teaser}
\end{teaserfigure}

\received{20 February 2007}
\received[revised]{12 March 2009}
\received[accepted]{5 June 2009}

\maketitle

\section{Introduction} \label{sec:intro}

%GNNUEM: bipartite Graph Explanation and Mitigation
% Social problem and motivation of the work: unfairness in recommender systems, access to every demographic group to the same quality of service, provide users with "control" of their recommendations explaining them how a certain action could affect recommendations (i.e. improve fairness if they choose an item over other ones).
Novel recommender systems have become more and more effective and sophisticated, excelling in satisfying the users' preferences.
The complexity of novel systems functioning has also increased dramatically, laying the groundwork for worrying questions~\cite{DBLP:conf/wsdm/GhazimatinBRW20}.
%Recent work  addressed this issue by providing an explanation in the form of 
%\cite{BALLOCCU2022100404} NEEDED TO TALK ABOUT DIFFERENCES BETWEEN POST-HOC EXPLANATIONS AND SELF-EXPLAINABLE
Such issue comes in addition with the prominent importance of preserving properties that go beyond recommendation effectiveness, such as trustworthiness~\cite{WangZWLR22}, fairness~\cite{WanngMZLM22}, and explainability~\cite{ZhangC20}. 
However, all these issues are usually treated by the modern literature as independent perspectives, mostly tackled one at a time.
Taking as an example algorithmic fairness (which is also the main case study in our work), while it is of uttermost importance to provide the end users and the content providers with equitable recommendations, it is also important for service providers (e.g., an online platform) to understand {\em why} the model behind their platform is unfair.
Hence, tackling algorithmic fairness in an explainable way is a central yet under-explored area.
The few existing approaches that explain unfairness in recommendation exploit user and item features to characterize the disparities generated by a model \cite{DBLP:conf/sigir/GeTZXL0FGLZ22,DBLP:journals/ipm/DeldjooBN21}.
Though their methods successfully uncover the features related to the model disparities, the generated explanations are limited to those selected features.
The latter might be challenging to obtain, given that most recommendation models work with user-item interaction data.

Shifting the focus to other areas within explainable artificial intelligence, various techniques have been employed to determine the relevant data entities that can serve as explanations for diverse tasks.
Counterfactual methods have recently emerged as an effective way to explain the predictions produced by models based on Graph neural networks (GNNs)~\cite{hamilton2017gnn,kipf2017gcn,velivckovic2017gat,zhang2022gnn-book}.
GNNs have proven to be effective in modeling graph data in several domains, such as information retrieval~\cite{cui2022can}, recommender systems~\cite{DBLP:conf/sigir/0001DWLZ020,ying2018gnn-recsys}, natural language processing~\cite{yao2019gnn-nlp} and user profiling~\cite{Chen2021-CatGCN,chen2019gnn-userprofiling,Yan2021-RHGN}.
Approaches driven by counterfactual reasoning have also been used to guarantee algorithmic fairness in GNN-based models by manipulating the topological structure~\cite{DBLP:conf/uai/AgarwalLZ21,DBLP:conf/wsdm/MaGWYZL22,DBLP:conf/www/WangLLW22} under various downstream tasks.
However, to the best of our knowledge, no approach was ever proposed to explain unfairness in GNN-based recommender systems.
Filling this research gap goes beyond a simple application of counterfactual explanations methods for GNNs, so as to uncover unfairness in recommender systems.
Indeed, the original methods to explain the predictions in GNN-based models are applied to classic graphs~\cite{DBLP:conf/aistats/LucicHTRS22,DBLP:conf/pkdd/KangLB21,DBLP:journals/corr/abs-2208-04222,DBLP:conf/kdd/YuanTHJ20}, while recommender systems are characterized by a bipartite nature, since they bridge the interactions between two types of entities (nodes), i.e., users and items.
This leads to efficiency issues, uncovered in detail in Section~\ref{sec:method}.

In this work, we propose a shift of paradigm, by presenting a framework, named {\tt GNNUERS} (\textbf{GNN}-based \textbf{U}nfairness \textbf{E}xplainer in \textbf{R}ecommender \textbf{S}ystems), which perturbs the original bipartite graph in order to identify a set of user-item interactions leading to user unfairness in GNN-based recommender systems.
Thanks to our approach, user unfairness can be uncovered by the selected edges, i.e. the user-item interactions.
Their removal from the original graph can result in fairer recommendations for the end users, thus globally explaining under which conditions a model generates disparities.
Concretely, our approach is guided by the demographic parity principle, which ensures that all demographic groups receive the same recommendation utility, {\color{black} where the utility of a recommendation denotes its usefulness for (a group of) users}.
Under this paradigm, we propose a perturbation mechanism driven by counterfactual reasoning.
The goal is to alter the users' recommendation by assuming a set of edges is missing from the graph.
In other words, the original bipartite graph is modified under the assumption that certain users have not interacted with the items associated with the selected edges (in the counterfactual world), while the graph includes such edges in reality (in the actual world).
The selection process of the user-item interactions is driven by a loss function that combines two terms: i) minimizing the disparity in recommendation utility across demographic groups, and ii) minimizing the number of perturbed edges.
Specifically, the user-item interactions are cherry-picked such that a model using a graph missing such interactions is able to generate fairer recommendations during the inference phase.

It should be noted that the loss function guiding the perturbation mechanism has access to the ground truth labels of the subset of data used for evaluation.
It follows that our approach is able to monitor the recommendation utility measured on the evaluation set for all the demographic groups.
This aspect is crucial to select the specific user-item interactions that lead to the utility disparity across demographic groups in the recommendations generated by GNN-based systems.
By leveraging the ground truth labels of the evaluation subset of data, \texttt{GNNUERS} distinguishes itself from other recently proposed procedures in the recommendation literature.
While these procedures aim at modifying the recommendations to mitigate the unfairness issue, our method focuses on explaining it instead.
Therefore, \texttt{GNNUERS} can generate explanations of user unfairness at the model level, also denoted as global explanations, that can support system designers and service providers with insights on how to improve the model fairness.
This claim is supported by the experiments in Section~\ref{sec:eval}, which prove our framework is able to systematically explain user unfairness, and also to uncover existing patterns that justify why unfairness was characterized.

Our contributions can be summarized as follows:
\begin{itemize}
    \item Based on the research gaps existing in the current literature (Section~\ref{sec:related}), we formulate the problem of explaining unfairness in GNN-based recommender systems (Section~\ref{sec:prelim}) and propose a framework to generate global counterfactual explanations of unfairness (Section~\ref{sec:method});
    \item We validate our proposal on three state-of-the-art models under four data sets, measuring key utility metrics (Section~\ref{sec:eval}), and finally discussing the implications of our work for future advances in this research area (Section~\ref{sec:discuss}).
\end{itemize}

\section{Related Work} \label{sec:related}

Our study has the primary goal of explaining unfairness in recommender systems based on graph neural networks, which is a naturally multidisciplinary topic.
Therefore, we first contextualize our study with respect to prior work that aimed at describing and explaining global issues related to the whole system (e.g., unfairness, instability) in recommendation or on graphs, and not the single model predictions.
We then outline the efforts directed by researchers towards explainability and unfairness mitigation in recommendation.
Finally, we provide an overview of existing methods adopting counterfactual explanations in graph neural networks.

\subsection{Beyond Predictions Explanation in Recommendation and on Graphs} \label{subsec:beyond_predic} % CASPER non usa GNN, usa un grafo di interazioni

First attempts to explain aspects beyond predictions in recommendation, e.g., unfairness, were made in \cite{DBLP:journals/ipm/DeldjooBN21,DBLP:conf/sigir/GeTZXL0FGLZ22}.
\cite{DBLP:journals/ipm/DeldjooBN21} aimed at estimating the dependency between data characteristics and unfairness, \cite{DBLP:conf/sigir/GeTZXL0FGLZ22} sought to finding the minimal perturbation to user and item features that can reduce exposure unfairness.
It should be noted that these explanations are therefore based on pre-computed features and their importance with respect to the unfairness.
Conversely, \texttt{GNNUERS} touches on user-item interactions, i.e. the main source of personalization for collaborative filtering models, including the GNN-based ones.
While we fully acknowledge the fact that these two aspects are complementary, user-item interactions are generally present in any dataset adopted in a recommendation scenario, whereas user and item features (e.g., those in \cite{DBLP:conf/sigir/GeTZXL0FGLZ22}) might not always be extracted from the available data.
Furthermore, the explanatory framework used by \cite{DBLP:journals/ipm/DeldjooBN21} requires a significant amount of simulations, impractical for models with significant training times, and the discovered patterns are informative only twith respect to the limited collection of selected data characteristics.

Unfairness was also explained in GNNs trained for node classification tasks \cite{DBLP:conf/aaai/DongW0LL23}.
The authors devised an efficient method to quantify to what extent fairness is impacted by the removal of a single node from the training set without re-training the GNN.
In a recommendation scenario, such technique could label some users or items as harmful, describing unfairness as their mere presence in the graph.
Conversely, \texttt{GNNUERS} addresses unfairness in the more complex recommendation task by detecting the users' behavior (user-item interactions) causing biased recommendations.

Other researchers have investigated to what extent recommender systems are sensitive to minimal input data perturbations.
For instance, the method proposed in \cite{DBLP:conf/cikm/OhUMK22} aims to find the minimal perturbation that causes the highest instability in the recommendations, by analyzing the effect of a perturbation over the interaction graph in a black-box setting.
However, such method is tailored to decrease recommendation stability and might be used to merely explain such criterion.
Conversely, \texttt{GNNUERS} can be adopted to optimize different objectives, including unfairness explanation and recommendation instability.
For the sake of scope, our study in this paper explores the former and leaves the latter as a future work.

\subsection{Explainability and User Fairness in Recommendation}

Recent work in the recommendation field put much emphasis on improving explainability in recommendation \cite{DBLP:journals/corr/abs-2202-06466} by leveraging post-hoc techniques \cite{DBLP:journals/is/BrunotCCLV22,DBLP:conf/sac/ZhongN22}, explanation paths \cite{DBLP:conf/sigir/BalloccuBFM22,DBLP:conf/www/Zhang0SAZFS23}, counterfactual reasoning \cite{DBLP:conf/www/GhazimatinPRW21,DBLP:journals/corr/abs-2208-04222,DBLP:conf/cikm/TanXG00Z21}.
Nonetheless, these methods are specifically designed to generate instance-level explanations \cite{DBLP:journals/pami/YuanYGJ23}, that is each of them describes why an item was recommended to a user.
On the other hand, \texttt{GNNUERS} provides explanations at the model level~\cite{DBLP:journals/pami/YuanYGJ23} due to unfairness being a concern within the scope of the entire recommender system, rather than within the scope of individual user-item interactions.
{\color{black} In more detail, a model-level explanation could benefit the system designers to understand how the system affects the user experience and how it could be improved.
However, a global explanation would not be informative at the level of an individual user.
Indeed, a user could not contextualize the explanation with their personal experience, given that a global explanation is not personalized as local ones.}
Our method is also different from the mitigation methods of the unfairness caused by instance-level explanations~\cite{DBLP:journals/corr/abs-2210-15500}, where the goal is to guarantee the explanations are not biased, e.g., with respect to the generated text.

Also the attention to issues of unfairness, from both the user and provider perspectives, has been increasing in recommendation.
Most works focusing on the end users have modeled fairness at a group level, with a primary focus on gender- and age-based demographic groups, and often accompanied unfairness assessments with technical contributions for mitigating it \cite{10.1007/978-3-030-99736-6_37,DBLP:journals/ftir/Ekstrand0B022,DBLP:conf/fat/EkstrandTAEAMP18,DBLP:conf/fat/BurkeSO18,DBLP:conf/fedcsis/FrischLG21,DBLP:conf/www/LiCFGZ21,DBLP:conf/sigir/LiCXGZ21,DBLP:conf/aaai/WuWWH021,10.1145/3564285}.
These methods however do not consider algorithmic ways to explain the causes behind the detected disparities, letting mitigation methods attempt to reduce such disparities by optimizing a certain loss function.
Although these mitigation methods usually led to a decrease in unfairness, the underlying causes of unfairness remain unclear, consequently preventing researchers and service providers from devising more conceptually informed mitigation methods. 
We particularly observed that it is generally hard to link the mitigation method logic to the underlying aspects causing unfairness in a given domain. 
Compared to these prior works, \texttt{GNNUERS} not only aims to identify prior user interactions that potentially led the model to provide unfair recommendations, but also investigates the structural characteristics of the perturbed interactions.
It follows that \texttt{GNNUERS} makes it possible to derive conceptual insights that support a better understanding of the unfairness phenomenon, working to the advantage of system designers and service providers. 

\subsection{{\color{black} Explainability} in GNNs}

GNNs can be directly applied to graphs to provide an easy way to do node-level, edge-level, and graph-level prediction tasks. 
A notable advantage of GNNs is that they can capture the dependence of graphs via message passing between the graph nodes. 
Unlike standard neural networks, GNNs retain a state representing information from their neighbourhood with arbitrary depth.

Yet, the methods addressing explainability about unfairness in GNNs are limited \cite{DBLP:conf/aaai/DongW0LL23} (described in Section \ref{subsec:beyond_predic}).
On the other hand, several prior works have proposed to mitigate unfairness in GNNs \cite{DBLP:conf/uai/AgarwalLZ21,DBLP:conf/wsdm/MaGWYZL22,DBLP:conf/www/WangLLW22}, specifically by manipulating the graph topological structure.
Such methods have never been investigated on bipartite graphs adopted for recommendation, and their operationalization of unfairness and their mitigation operation could not be extended to explain the issue, as required in our case.
{\color{black} Meanwhile, model-level explainability methods for GNNs~\cite{DBLP:conf/kdd/YuanTHJ20,DBLP:conf/wsdm/HuangKMRS23} have been limited to generate graphs that explain how the models globally behave in graph classification tasks.
However, the classified graphs are much smaller than user-item interaction networks adopted in recommendation, and the graph generation process iteratively applies a single modification.
This operation prevents the existing methods from scaling to datasets commonly used in recommendation.}

{\color{black} A relevant amount of works were devoted to leveraging counterfactual techniques to explain GNN predictions}~\cite{DBLP:conf/aistats/LucicHTRS22,DBLP:conf/pkdd/KangLB21,DBLP:journals/corr/abs-2208-04222,DBLP:conf/wsdm/HuangKMRS23}.
These techniques aim to find the minimal perturbation to the input (graph) data that makes the GNN change its predictions.
The goal of such approaches is related to adversarial attacks, but the latter aim at degrading the model performance instead of generating informative explanations \cite{DBLP:conf/aistats/LucicHTRS22}.
Nevertheless, none of them focused on explaining fairness, but merely addressed explanations of individual predictions.
Differently from them, \texttt{GNNUERS} leverages counterfactual techniques to find the minimal perturbation to the input (graph) data such that the unfairness of the recommendations produced by the GNN-based model is reduced. 
Our method therefore differs from other existing solutions from several perspectives.
First, we target GNNs applied to bipartite graphs, instead of focusing on more general graphs typically considered in the machine learning field. 
Second, we adopted a different notion of counterfactuality, which does not only require that the prediction changes, but also that predictions lead to a certain property.
Third, we go beyond the mere creation of the counterfactual explanations at the instance level, and investigate structural properties of the perturbed data at the model level.

\section{Problem Formulation} \label{sec:prelim}

Our paper aims to explain user unfairness in recommendations generated by graph neural network models. 
Therefore, we first describe the recommendation scenario from a graph perspective. 
We then formulate the target task, namely user unfairness explanation.
Finally, we introduce the definition of fairness adopted in our study.

\subsection{Recommendation Task}

In recommendation, the goal of the preference model is typically predicting whether or to what extent an (unseen) item would potentially be of interest for a user.
In a common scenario, the model uses past interactions between two main entities, namely users $U$ and items $I$, to learn preference patterns. 
Each user $u \in U$ is assumed to have interacted with a certain item $i \in I$ in case they rated, liked, or clicked on such item, depending on the applicative scenario. 
The set of items $I_u$ a user interacted with is referred to as the $u$'s history.

Graphs are structures that represent a set of entities (nodes) and their relations (edges). 
GNNs operate on graphs to produce representations that can be used in downstream tasks. 
In our case, user-item interactions can be represented by means of an undirected bipartite graph $\mathcal{G} = (U, I, E)$, where $E$ is the set of edges representing the interactions and $U \cup I$, with $n = |U| + |I|$, is the set of nodes. 
No edge exists between nodes of the same type, i.e. $E = \{(u, i)\ | \; u \in U, i \in I\}$.
The recommendation problem can be then solved by leveraging GNNs on a linking prediction task to predict potentially interesting links between users $U$ and items $I$ in the bipartite graph $\mathcal{G}$.

Let $f(A; W) \rightarrow \hat{R}$ be any GNN, where $A$ is an $n \times n$ adjacency matrix representing $\mathcal{G}$, $W$ is the learned weight matrix of $f$, and $\hat{R}$ is a $|U| \times |I|$ user-item relevance matrix, with $\hat{R}_{u,i}$ being the linking probability between user $u$ and item $i$.
In other words, $A$ is the input of $f$, and $f$ is parameterized by $W$.
$f$ predicts the user-item relevance matrix $\hat{R}$ by combining the normalized adjacency matrix $L = D^{-\frac{1}{2}} A D^{-\frac{1}{2}}$, where $D_{j,j} = \sum_k{A_{j,k}}$ are entries in the degree matrix $D$, and $W$ is learnt according to the GNN implementation.
Given $\hat{R}$ and a user $u$, items in $I$ are sorted based on their decreasing relevance in $\hat{R}_u$, and the top-$k$ items are recommended to user $u$.
We refer to the list of items recommended to user $u$ as $Q_u$ and to the set of all recommended lists as $Q$.  

\subsection{Unfairness Explanation Task} \label{subsec:probform}

An unfair algorithm is one whose decisions are skewed toward a particular group of people \cite{DBLP:journals/csur/MehrabiMSLG21}.
In light of this, it follows that an unfairness explanation process cannot be conducted for a single individual.
Hence, we aim to explain why a recommender system is unfair at the model level \cite{DBLP:journals/pami/YuanYGJ23} to support designers and service providers with insights on how to improve the model fairness.
To do so, we decided to adopt counterfactual reasoning techniques \cite{DBLP:conf/aistats/LucicHTRS22,DBLP:conf/pkdd/KangLB21}. 
In our context, we assume to model counterfactual explanations according to the users' history {\color{black} and provide a formal definition.

\begin{definition}[Counterfactual Explanation of Recommendation Unfairness in GNNs] \label{def:cf_exp_unfair}
    Let $E$ be a set of user-item interactions represented as edges in an adjacency matrix $A$. Let $f$ be a trained GNN. Let $\Phi(\cdot) \in [0, 1]$ be a fairness metric adopted on a set of recommendations, where $\Phi(f(A)) = 0$ denotes perfect fairness. Let the perturbation (e.g., deletion) of the edges in $\tilde{E} \subseteq E$ result in the adjacency matrix $\tilde{A}$. $\tilde{E}$ is a counterfactual explanation of the recommendation unfairness if $\Phi(f(\tilde{A})) < \Phi(f(A))$.
\end{definition}

Definition~\ref{def:cf_exp_unfair} introduces how to identify one of the possible sources of the unfairness as a subset of edges included in $A$.
The presence of the subset of edges $\tilde{E}$ in $A$ leads a GNN to generate biased recommendations that advantage a specific group of users.
Hence, perturbing $\tilde{E}$ could make the system fairer, which follows the logic of existing explainability techniques for provider fairness~\cite{DBLP:conf/sigir/GeTZXL0FGLZ22}.

However, Definition~\ref{def:cf_exp_unfair} does not address the amount of perturbed edges and the extent to which the unfairness level was reduced.
It follows that several explanations could be identified as long as the recommendations generated by the GNN with the perturbed graph are fairer, even if such explanations are not informative enough of the recommendation unfairness.
Therefore, we formulate the problem of identifying an \textit{Optimal Counterfactual Explanation}.

\begin{problem}[Identification of an Optimal Counterfactual Explanation of Recommendation Unfairness in GNNs] \label{prob:identify_unfair_exp}
    Let $f$ be a GNN, $E$ be a set of edges represented in an adjacency matrix $A$, and $\Phi(f(A))$ be the fairness level of the recommendations generated by $f$ based on $A$. Our goal is to identify a counterfactual explanation $\tilde{E}$ (with corresponding adjacency matrix $\tilde{A}$) that leads to perfect fairness, i.e. $\Phi(f(\tilde{A})) = 0$, by means of the lowest number of perturbations, i.e. $|\tilde{E}|$ is minimal.
\end{problem}

It follows that Problem \ref{prob:identify_unfair_exp} could be solved by identifying the minimal subset of user-item interactions that make the recommendations perfectly fair when such interactions are removed from the adjacency matrix.
We denote such interactions as an \textit{Optimal Counterfactual Explanation}, which is formally defined as follows:

\begin{definition}[Optimal Counterfactual Explanation of Recommendation Unfairness in GNNs] \label{def:opt_cf_exp_unfair}
    Let $\tilde{\mathcal{E}} = \{\tilde{E}_1, \tilde{E}_2, \dots\}$ be a set of counterfactual explanations of the unfairness estimated on the recommendations generated by a GNN $f$. According to Definition \ref{def:cf_exp_unfair}, $\Phi(f(\tilde{A})) < \Phi(f(A))$ is true $\forall \tilde{E} \in \tilde{\mathcal{E}}$ used to perturb $A$ and generate $\tilde{A}$. We define a counterfactual explanation $\tilde{E}_j$ as optimal if:
    \begin{displaymath}
        \begin{aligned}
            1. \quad & \Phi(f(\tilde{A}_j)) = 0 \\
            2. \quad & |\tilde{E}_j| \leq |\tilde{E}_k|, \forall E_k \in \tilde{\mathcal{E}}, j \neq k \\
            3. \quad & \text{utility}(f(\tilde{A}_j)) \approx \text{utility}(f(A))
        \end{aligned}
    \end{displaymath}
    where $\tilde{A}_j$ is the adjacency matrix generated by perturbing $\tilde{E}_j$ on $A$, $\text{utility}(\cdot)$ is a recommendation utility metric.
\end{definition}

Definition~\ref{def:opt_cf_exp_unfair} adds a requirement to the previous discussion, stating that the recommendation utility before and after the perturbation should be preserved as much as possible.
This requirement is essential to avoid scenarios that do not reflect a proper unfairness explanation.
For instance, if no user received at least a relevant item, the recommendations would not exhibit any disparate treatment, but the system would not be useful in practice.

We then remark our goal to identify an optimal counterfactual explanation of recommendation unfairness, but underline that Definition~\ref{def:opt_cf_exp_unfair} does not claim such explanation is granted to exist.
Therefore, our goal is approximated to find a counterfactual explanation that is as close as possible to the optimal one.
This involves perturbing the minimal subset of edges that result in an unfairness level as close as possible to zero and an utility level as close as possible to the original one.
}

Algorithmic unfairness has been operationalized through numerous notions, often dependent on the context and the application \cite{DBLP:conf/kdd/SinghJ18}. 
It follows that there is no consensus in the recommender systems community on a gold standard definition to apply.
Motivated by its increasingly recognized importance in prior work in top-$k$ recommendation \cite{10.1007/978-3-030-99736-6_37, 10.1145/3564285}, we decided to model fairness according to the notion of demographic parity.
% , denoted with the function $dp(Q) : Q \rightarrow \mathbb{R}_{\geq 0}$ (the closer to 0 the fairer).
Nonetheless, our formulation and method is flexible to accommodate other notions of fairness.
In the context of recommendations, a model meets demographic parity when the recommendation utility estimates across demographic groups are not systematically different. 
{\color{black} Under this demographic parity notion of fairness, our goal is to generate recommendation lists with closer utility estimates across demographic groups than the original recommendations $Q$.
We pursue such goal by generating a perturbed adjacency matrix $\tilde{A}$ that results in fairer recommendations when a trained GNN uses $\tilde{A}$ instead of $A$ during inference. 
We also constrain the number of perturbed edges with respect to the original adjacency matrix $A$.}
% Under this demographic parity notion of fairness, our goal is to generate a perturbed adjacency matrix $\tilde{A}$ that modifies the predictions of a trained GNN, resulting in recommended lists with closer utility estimates across demographic groups than the original recommendations $Q$, constrained to the number of perturbed edges with respect to the original adjacency matrix $A$.
Formally, we seek to minimize the following objective function:

\begin{equation} \label{eq:loss}
    \mathcal{L}(A, \tilde{A}) = \mathcal{L}_{fair}(A, f(\tilde{A}; W)) + \mathcal{L}_{dist}(A, \tilde{A})
    \vspace{2mm}
\end{equation}

where $\mathcal{L}_{fair}$ is the term monitoring fairness, operationalized according to the notion of demographic parity, $\mathcal{L}_{dist}$ is the term controlling the distance between the perturbed adjacency matrix $\tilde{A}$ and the original one $A$. %, $\tilde{f}$ is a modified GNN that uses $\tilde{A}$ instead of $A$ and it will be described in detail in Section \ref{subsec:graphgen}.
In the next section, we describe the way $\tilde{A}$ is generated and how the objective function is translated into a loss function to be minimized with our approach.

\section{Methodology: GNNUERS} \label{sec:method}

In this section we present \texttt{GNNUERS}, a method able to explain unfairness in graph-based recommenders, solving the problem introduced in Section \ref{subsec:probform}. We introduce the method by its main components: $(a)$ first, the perturbation mechanism of the bipartite graph, which allows to alter the interactions between users and items in a differentiable way, and $(b)$ the two-term loss function that guide the selection of the edges to be perturbed.
%, (iii) the mitigation procedure that leverages the generated counterfactual explanations to mitigate the unfairness of the model recommendations.

\subsection{Bipartite Graph Perturbation} \label{subsec:perturb}
%Several works benefit from the perturbation of network's topological structure to solve a specific problem: guaranteeing counterfactual fairness in node classification \cite{DBLP:conf/wsdm/MaGWYZL22}, explaining predictions in node classification \cite{DBLP:conf/aistats/LucicHTRS22} and link prediction \cite{DBLP:conf/pkdd/KangLB21}.

% Present the way the matrix P perturbs the original one. It must be precised how we operate, different from Silvestri et al. because only the accounted entries are modified, and no operation is performed on the original graph. For edge addition => only the entries with zero are modified, For edge deletions => the original Graph is substituted with P, which is populated only on the entries of the original Graph (the binary version is identical to the original Graph)
Our graph perturbation approach is inspired by previous work on GNNs explainability for binary classification on plain graphs \cite{DBLP:conf/aistats/LucicHTRS22}.
However, since \texttt{GNNUERS} aims to perturb a bipartite graph generated for recommender systems, it presents several differences.
In \cite{DBLP:conf/aistats/LucicHTRS22}, a perturbation matrix $P$ is populated to then generate the perturbed matrix $\tilde{A} = P \odot A$\footnote{$\odot$ denotes the Hadamard product.}. 
Optimizing for $P$ can eventually include indices for zero entries in $A$.
While for plain graphs this method results to be efficient, for bipartite graphs it can be memory inefficient, mainly because it requires to store a perturbation value also for the user-user and item-item links\footnote{\color{black} Given our focus on explaining unfairness by means of user behavior, we decided not to consider user-user and item-item links in \texttt{GNNUERS}. We leave this interesting perspective to its future extensions.}, not present in bipartite graph per definitionem.
To overcome these limitations, the perturbation in \texttt{GNNUERS} is {\color{black} driven by a vector $p \in \mathbb{N}^{|\tilde{E}|}$, where $\tilde{E} \subseteq E$ is the subset of edges selected for perturbation (as in Definition \ref{def:cf_exp_unfair}-\ref{def:opt_cf_exp_unfair}).
In our experiments, we consider all the edges of the training graph in the \texttt{GNNUERS} perturbation process, i.e. $\tilde{E}$ includes all the pairs $(u, i)$ that correspond to non-zero entries of $A$.}
Our method is memory efficient, especially under sparse graphs, since it needs to store perturbation values only for the non-zero entries of $A$.

% Since it is bipartite, \tilde{A} is generated from a vector p with a number of entries equals to the entries that can be modified (only entries user-item, because user-user and item-item makes no sense in our protocol). The indices to be modified are pre-computed and used to populate \tilde{A}. The same thresholding is used to generate the binary version.

Given our unfairness explanation task, we derive a perturbed matrix $\tilde{A}$, resulting in fairer recommendations when a trained GNN uses $\tilde{A}$ instead of $A$ during the inference phase.
{\color{black} The mechanism that controls how the vector $p$ perturbs the edges included in $\tilde{E}$ is determined by the relation $H = \{(p_j, \tilde{e}_j) \; | \; 1 \leq j \leq |\tilde{E}|, \tilde{e}_j \in \tilde{E}\}$.
In other words, an edge $\tilde{e}_j$, i.e. a distinct non-zero entry $A_{u,i}$, is associated with a specific entry $p_j$.
An entry $p_j = 0$ denotes the edge $A_{u, i}$ (i.e. $\tilde{e}_j$) is deleted in the perturbed adjacency matrix, i.e. $\tilde{A}_{u, i} = 0$.
Thus, the perturbed matrix $\tilde{A}$ is populated as follows:
\begin{equation} \label{eq:perturb}
    \begin{aligned}
        \quad & \tilde{A}_{u,i} =
        \begin{cases}
            {p}_{j} & if \: (p_j, \tilde{e}_j) \in H, \tilde{e}_j = (u,i) \\
            A_{u,i} & otherwise
        \end{cases} \\
    \end{aligned}
\end{equation}
It should be noted that the way the relation $H$ is defined does not affect the perturbation process.
In particular, the perturbation is invariant of the order defined for the edges in $\tilde{E}$ used to construct $H$.
}

Following \cite{DBLP:conf/aistats/LucicHTRS22, DBLP:conf/cvpr/SrinivasSB17}, we generate $p$ through an optimization process that leverages a real valued vector $\hat{p}$.
Once optimized, we apply a sigmoid transformation.
Then, entries are binarized such that values {\color{black} greater or equal than} $0.5$ become 1, while values {\color{black} smaller than} $0.5$ become 0, obtaining eventually $p$.
The initialization of $\hat{p}$ should guarantee $\tilde{A} = A$, i.e. a real-valued $\alpha$ is selected to initialize $\hat{p}$, such that $p_j = 1, \forall j \in [0, B)$. In all the experiments in Section \ref{sec:eval} we set $\alpha = 0$, leaving the analysis of other initialization values as a future work.
%Note that the initialization of $\hat{p}$ should guarantee $\tilde{A} = A$, regardless of the selected perturbation task. Therefore, for a given real-valued $\alpha$, the initialization for each task should be performed as follows:
% \begin{enumerate}[label=(\alph*)]
%     \item \textbf{edge deletion}: $\hat{p}_i = \alpha$ such that $p_i \ge 1$
%     \item \textbf{edge addition}: $\hat{p}_i = \alpha$ such that $p_i \le 0$
% \end{enumerate}

\subsection{Perturbed Graph Generation} \label{subsec:graphgen}

Based on the protocol described above, \texttt{GNNUERS} modifies the adjacency matrix edges by means of the perturbation vector $p$.
The decision process of which edges will be deleted is performed by the counterfactual model $\tilde{f}$. $\tilde{f}(A, W; \hat{p}) \rightarrow \tilde{R}$ extends the GNN-based recommender system $f$ using $\hat{p}$ as parameter and the frozen weights $W$ learnt by $f$ as additional input.
% In detail, $\tilde{f}$, similarly to $f$, predicts the altered relevance matrix $\tilde{R}$ by combining the normalized version of the perturbed adjacency matrix $\tilde{L} = \tilde{D}^{-\frac{1}{2}} \tilde{A} \tilde{D}^{-\frac{1}{2}}$, where $\tilde{D}_{u,u} = \sum_i{\tilde{A}_{u,i}}$, with $W$ according to the implementation of the original model $f$.
{\color{black} $\tilde{f}$ predicts the altered relevance matrix $\tilde{R}$ similarly to $f$.
In detail,} $\tilde{f}$ combines the normalized version of the perturbed adjacency matrix $\tilde{L} = \tilde{D}^{-\frac{1}{2}} \tilde{A} \tilde{D}^{-\frac{1}{2}}$, where $\tilde{D}_{u,u} = \sum_i{\tilde{A}_{u,i}}$, with $W$ according to the implementation of the original model $f$.
Therefore, $\tilde{f}$ learns only $\hat{p}$, while the weights $W$, already optimized by $f$ to maximize the recommendation utility, remain constant.

As explained in Section \ref{subsec:perturb}, $p$ is generated from $\hat{p}$, whose values get updated during the learning process. At different steps of the latter,
the values of $\hat{p}$ could oscillate close to the threshold that determines if $p$ will be 0 or 1 at the respective indices.
Considering aspects such as floating-point errors or dropout layers, the oscillation could negatively affect the update of $\hat{p}$, due to previously perturbed edges being restored, or vice versa.
To counter this phenomenon, the perturbation algorithm is constrained by the usage of a policy that prevents a deleted edge from being restored, such that the number of perturbed edges follows a monotonic trend.
%The usage of such policy will be considered as the base explanation algorithm.

{\color{black} In order to properly explain unfairness, the edges removed by \texttt{GNNUERS} should maintain the recommendation properties other than unfairness as similar as possible as they were before perturbing the graph.
In light of this perspective, a deletion cost $\lambda$ could be defined for each edge $\tilde{e} \in \tilde{E}$, such that deleting an edge could have a greater impact over deleting another edge.
For instance, the cost of an edge should be high if removing it would significantly alter the overall recommendation utility.
In this version, \texttt{GNNUERS} does not apply any cost to edges, which corresponds to assuming that all edges have the same deletion cost.
The cost of removing an edge is implicitly determined in a data-driven way by the model during the optimization process.
We leave the analysis of explicit costs as a future work.}

\subsection{Loss Function Optimization} \label{subsec:lossoptim}

% Describe the loss that we want to optimize. L_pred can be any function is desired and the setting is similar to the one of Silvestri et al. We do not use an indicator function because we are not interested in finding the minimal perturbation to change the recommended lists, but the perturbation that lead to fairer outcomes.
% We then used the fairness loss function taken from "A Multi-objective Optimization Framework for Multi-stakeholder Fairness-aware Recommendation". We used it in its simple binary form => demographic parity between demographic groups (only age and gender, by no means binary, but used to compare the results with other mitigation methods).

The previous section introduced $\tilde{f}$, the counterfactual model responsible of the generation of the perturbed adjacency matrix $\tilde{A}$.
The optimization of the perturbation vector $\hat{p}$ is guided by the loss function defined in \eqref{eq:loss}, where $\mathcal{L}_{fair}$ is based on Demographic Parity (DP), the fairness notion described in Section \ref{subsec:probform}.
Recent works \cite{10.1007/978-3-030-99736-6_37,10.1145/3564285} operationalized DP as the mean of the absolute pair-wise utility difference across all demographic groups. Formally:
\begin{equation} \label{eq:fairloss}
    \mathcal{L}_{fair}(A, \tilde{R}) = \frac{1}{\binom{|G|}{2}}\sum_{1 \leq i < j \leq |G|}{\left\Vert S(\tilde{R}^{G_i}, A^{G_i}) - S(\tilde{R}^{G_j}, A^{G_j}) \right\Vert^2_2}
\end{equation}
where $S$ is a recommendation utility metric, $G$ is the set of considered demographic groups, $\tilde{R}^{G_i}$ ($\tilde{R}^{G_j}$) and $A^{G_i}$ ($A^{G_j}$) denote, respectively, the altered relevance sub-matrix and the adjacency sub-matrix with respect to the users in $U^{G_i}$ ($U^{G_j}$).
{\color{black}Thus, $\mathcal{L}_{fair}$ is equivalent to $\Phi$ in Definition~\ref{def:cf_exp_unfair}-\ref{def:opt_cf_exp_unfair}.}

% Introduce the choice of using binary setting. From now on, we will denote the groups as protected and unprotected for the ones with lower and higher utility respectively. A toy example would be helpful to depict the goal of the fairness loss
Following works that proposed methods to mitigate or explain unfairness in recommendation \cite{DBLP:conf/www/LiCFGZ21,DBLP:journals/ipm/AshokanH21,DBLP:conf/fat/KamishimaAAS18,DBLP:conf/sigir/GeTZXL0FGLZ22}, we focus on a binary setting, with sensitive attributes comprised of two demographic groups. %as in several datasets, e.g. gender in ML-1M.
For instance, given $G=\{males (M), females (F)\}$, $\mathcal{L}_{fair}$ aims at minimizing the utility disparity between males and females, with the optimal result being:
\begin{displaymath}
    \left\Vert S(\tilde{R}^{M}, A^{M}) - S(\tilde{R}^{F}, A^{F}) \right\Vert^2_2 = 0
\end{displaymath}
We denote the group with higher (lower) utility on the evaluation set as \emph{unprotected} (\emph{protected}).
This enables the reader to better contextualize the approach with respect to fairness.

% Then describe the loss that we used to measure NDCG, cite the paper that presented and explain how the data used to optimize (validation, train) is managed in the approximated loss function.
{\color{black} In our experiments, the recommendation utility metric $S$ was monitored through the Normalized Discounted Cumulative Gain (NDCG)~\cite{DBLP:journals/tois/JarvelinK02}. This metric measures the utility of a recommended list based on the position where the relevant items are placed.}
However, due to the non-differentiability of the sorting operation performed to compute NDCG, we adopt an approximated version~\cite{10.1145/3564285, DBLP:journals/ir/QinLL10}, which we refer as \emph{NDCGApproxLoss}:
\begin{equation}
    \begin{aligned} \label{eq:ndcgapprox}
        NDCGApproxLoss(r, a) = & \: -\frac{1}{DCG(a,a)}\sum_{i}{\frac{2^{a_i} - 1}{log_2(1 + z_i)}} \\
        \text{s.t.} \quad z_i = & \: 1 + \sum_{j \neq i}{\sigma \left(\frac{r_j - r_i}{\gamma}\right)}
    \end{aligned}
\end{equation}
where DCG is the Discounted Cumulative Gain, $r$ is the item relevance score produced by the recommender system, $\sigma$ is a sigmoid function, and $\gamma$ is a scaling constant. We fix $\gamma = 0.1$ for the experiments in Section \ref{sec:eval}, being the default value in the TensorFlow Ranking implementation.
\emph{NDCGApproxLoss} is adopted only when constructing the fairness objective in Equation \eqref{eq:fairloss}, while the original NDCG is used in the evaluation phase.
Given our unfairness explanation task, the ground truth labels used to measure \emph{NDCGApproxLoss} during the \texttt{GNNUERS} learning process are taken from the evaluation set.
Such approach is justified by the explanation task of the recommendation unfairness measured on the evaluation set itself, while for other tasks, e.g., mitigation ones, having access to the ground truth labels is a less realistic assumption \cite{DBLP:conf/sigir/RahmaniNDA22}.

% Describe L_dist as the number of edges that were modified in the original graph. Note that the graph is symmetrical, then 2 edges are 1 in reality, because the nodes are just swapped in the second one.
Any differentiable distance function can be adopted as the distance loss $\mathcal{L}_{dist}$ \cite{DBLP:conf/aistats/LucicHTRS22}. In \texttt{GNNUERS}, it is based on the absolute element-wise difference between $\tilde{A}$ and $A$, defined as follows:
\begin{equation}
    \mathcal{L}_{dist}(A, \tilde{A}) = \beta \frac{1}{2}\sigma\left(\sum_{i, j}{\left\Vert\tilde{A}_{i,j} - A_{i,j}\right\Vert^2_2}\right)
\end{equation}
%where $1/2$ is added due to the symmetrical structure of the bipartite adjacency matrix.
A sigmoid function is used to bound the distance loss to the same range of $\mathcal{L}_{fair}$, i.e. [0,1].
We used $\sigma(x) = |x| / (1 + |x|)$, which needs a higher number of perturbed edges to reach 1 compared with the popular logistic function, hence covering a wider range of values.
$\beta$ is a parameter that balances the two losses, due to the trend of $\mathcal{L}_{fair}$ to report values {\color{black} much smaller than} $0.5$, while $\mathcal{L}_{dist}$ gets rapidly close to 1 as more edges are perturbed.
We tested several values of $\beta$ in the range $[0.001, 2.0]$ and the best value for each model was selected for the experiments in Section \ref{sec:eval}.
%In all the experiments in Section \ref{sec:eval}, we set $\beta = 0.01$ as the value that works best to normalize $\mathcal{L}_{dist}$.

\subsection{Gradient Deactivation} \label{subsec:gradactive}

% Only one of the group is directly affected by the gradient. Depending on the downstream task (adding or deleting), the gradient for the unprotected or protected group respectively is deactivated. This does not imply that edges could not be deleted also for the other group, since the gradients affect also the items, which could modify the interactions of the deactivated group.
The optimization of \eqref{eq:fairloss} takes into account the approximate NDCG measured on the predicted recommendations for the protected and unprotected group.
The update of the real-valued perturbation vector $\hat{p}$ is then affected from the viewpoint of both demographic groups.
In particular, \texttt{GNNUERS} selects edges that could simultaneously optimize two objectives: increasing utility for the protected group and decreasing it for the unprotected one.
However, the edges that are going to be perturbed for one of the objectives could negatively affect the other one, and vice versa.
To this end, we perform a gradient \emph{deactivation} on the recommendations generated for the protected group, i.e. the back-propagation updates the perturbation vector only from the unprotected group viewpoint.
This procedure is applied only on the protected group, such that the \texttt{GNNUERS} objective is to delete the edges leading the unprotected group to enjoy a higher level of recommendation utility.

Deactivating the gradient does not limit the group of edges that can be perturbed because the optimization does not involve only the user nodes, but also the item ones.
Hence, \texttt{GNNUERS} could delete all the edges connected to an item node, both coming from user nodes of the unprotected and protected group.
For conciseness, we will use the terms \emph{deactivated} and \emph{activated} to characterize a group associated with inactive and active gradient respectively.

\subsection{Resources Usage} \label{subsec:resource_usage}

% The perturbation algorithm is O(B) where B is the number of entries to perturb. Considering the number of iterations to optimize the perturbed matrix, the total complexity is O(C * T), with C number of iterations.
% Memory footprint of the perturbation is optimized with sparsification
In this section, the two phases of the \texttt{GNNUERS} pipeline are examined in terms of memory footprint and execution time complexity.
The first phase regards the generation of the perturbed matrix $\tilde{A}$ at each step of the learning process by means of \eqref{eq:perturb}, which requires to store only the real-valued perturbation vector $\hat{p}$.
Leveraging a sparse representation of $A$ and $\tilde{A}$, the perturbation time complexity is dependent on the number of edges selected for perturbation $|\tilde{E}|$, i.e., $\mathcal{O}(|\tilde{E})|$.
The second phase, that is the optimization process directed towards learning $p$ (Sections \ref{subsec:graphgen}-\ref{subsec:lossoptim}), has no memory footprint and is executed for $C$ iterations.
{\color{black} The iterations timing is dependent on the cardinality of the user set and the number of user batches taken in input by \texttt{GNNUERS}}.
Hence, given {\color{black} $\Lambda$} the execution time for the inference step of the extended GNN-based recommender system $\tilde{f}$ and $\Psi$ the execution time of \eqref{eq:loss}, $\mathcal{O}({\color{black}\Lambda}\Psi C|\tilde{E}|)$ is the time complexity of the perturbed graph generation.

\section{Evaluation} \label{sec:eval}

In this section, we examine the \texttt{GNNUERS} explanations with experiments aimed at answering the following research questions:
\begin{itemize}
    % \item \textbf{RQ1}: To what extent is GNNUERS able to select counterfactual explanations that improve the operationalized fairness notion for the given ground truth labels?
    \item \textbf{RQ1}: Can the perturbation of the graph topological structure explain recommendation utility unfairness under the operationalised fairness notion?
    \item \textbf{RQ2}: Can the perturbation minimally affect the recommendation utility of the protected group while reducing the unprotected group one?
    %\item \textbf{RQ2}: To what extent the recommendations utility for each demographic group is affected by the edges perturbation?
    % \item \textbf{RQ3}: Does \texttt{GNNUERS} perturbation depend on the topological graph properties of the nodes associated to removed edges?
    \item \textbf{RQ3}: Can the categorization of user nodes through topological graph properties reveal the nature of unfairness?
    % \item \textbf{RQ4}: To what extent the recommendations lists generated on the perturbed graph differ from the original ones?
\end{itemize}

The data manipulation, training and assessment of the GNN-based recommender systems were built upon the framework Recbole \cite{DBLP:conf/cikm/ZhaoMHLCPLLWTMF21}. The experiments were ran on a A100 GPU machine with 80GB VRAM and 90GB RAM.

\subsection{Graph Topological Properties} \label{subsec:graph_properties}

\texttt{GNNUERS} identifies explanations in the form of user-item interactions that made a GNN-based recommender system generate unfair outcomes.
Each edge deleted from the graph unlinks a user and an item node, modifying the network topological structure and affecting the properties characterizing all the nodes, e.g., degree.
\texttt{GNNUERS} edges selection process can then be described by the properties of the nodes that constitute the removed edges.

To this end, we selected three properties that reflect different networks topological aspects and their relation to properties examined in recommender systems, e.g., popularity bias.
Let $z \in Z$ be a generic node of $\mathcal{G}$, i.e. $Z \subset U$ if $z$ is a user or $Z \subset V$ if $z$ is an item, the nodes properties are:
\begin{itemize}
    \item \textbf{Degree (DEG)}: the number of edges connected to each node. For a user node $u$ it represents the history length, i.e. $|I_u|$, for item nodes it represents their popularity.
    % \item \textbf{Sparsity (SP)}: it represents the tendency of a node to be connected to low-degree nodes. For user nodes it represents the tendency to interact with niche items, for item nodes it describes the disinterest of their peers, where users' disinterest is higher as their histories length is shorter. Formally:
    % \begin{equation} \label{eq:sparsity}
    %     SP_z = 1 - \frac{\sum_{i=1}^{|I_u|}{\frac{|\{z'' \; | \; (z', z'') \; \in \; E \; \land \; z' \; \in \; I_u \}|}{|Z|}}}{|I_u|}
    % \end{equation}
    \item \textbf{Density (DY)}: it represents the tendency of a node to be connected to high-degree nodes. For user nodes it represents the tendency to interact with popular items, for item nodes it describes the interest of their peers, where users' interest is greater as their histories length is longer. Formally, given $\mathcal{N}_z$ the neighbors set of a node $z$:
    \begin{equation} \label{eq:density}
        DY_z = \frac{\sum_{i=1}^{|\mathcal{N}_z|}{\frac{|\{z' \; | \; (\bar{z}_i, \; z') \; \in \; E \; \land \; \bar{z}_i \; \in \; \mathcal{N}_z \}|}{|Z|}}}{|\mathcal{N}_z|}
    \end{equation}
    \item \textbf{Intra-Group Distance (IGD)}: it represents how a node $z$ is close to the other nodes $z' \in Z/\{z\}$. Given the bipartite nature of recommender systems networks, we consider two users (items) being distant $n$ if the shortest path that connects them include $n$ items (users). IGD is the average of the number of nodes of the same type of $z$ normalized by their distance to $z$. Formally, given $N$ the graph diameter:
    \begin{equation} \label{eq:igd}
        IGD_z = \frac{\sum_{n=1}^{N}{\frac{|\{z' \; | \; \Gamma(z, \; z') \; = \; n\}|}{n}}}{|Z|}
    \end{equation}
    where $\Gamma$ measures the shortest path length between two nodes of the same type.
\end{itemize}

The selected properties can describe the context on which \texttt{GNNUERS} operates, i.e. the GNN-based recommender systems.
Moreover, insights on the unfairness can be uncovered by the variance of such properties across demographic groups, due to the intrinsic relationship between the given context and the unfairness.
In detail, the degree (DEG) of a user node represents the interest towards the available items and the amount of information that the GNN can leverage in the aggregation step, the density (DY) of a user node captures the inclination to engage with popular items, the intra-group distance (IGD) of a user node indicates the propensity to interact with items valued by fellow users within the same demographic group.
Moreover, the properties DEG and DY reflect the GNN ability to propagate information across the user nodes using the message passing mechanism, given that DEG and DY regard the nodes amount at the 1- and 2-hop distance respectively.

{\color{black}Our properties present similarities with assortativity mixing~\cite{PhysRevE.67.026126}, which estimates the propensity of nodes of the same group, e.g., females, to connect to each other.
This definition is mostly connected with IGD, given that it intrinsically leverages the groups information to discover how close fellow users are.
Nonetheless, assortativity mixing addresses only direct connections between nodes.
Conversely, IGD captures also the information about the distance among nodes of the same group, where such distance accounts for the bipartite nature of the network.}

\subsection{Data Preparation}

\begin{table}[!b]
\centering
\caption{\footnotesize Statistics of the four data sets used in our experimental protocol. \emph{Repr.} stands for \emph{Representation}, \emph{Min.} for \emph{Minimum}. \emph{G} stands for Gender, \emph{A} for Age, \emph{Gini} for the Gini coefficient applied over the values of the graph properties for each group. The line over the graph properties denotes their average.}
\label{tab:datasets}
\resizebox{.75\linewidth}{!}{
\begin{tabular}{rr|r|r|r|r}
\toprule
       &     &                   ML-1M~\cite{DBLP:journals/tiis/HarperK16} &                    FENG\footnote{\url{https://www.kaggle.com/datasets/chiranjivdas09/ta-feng-grocery-dataset}} &                  LFM-1K~\cite{DBLP:books/daglib/0025137} &                     INS\footnote{\url{https://www.kaggle.com/datasets/mrmorj/insurance-recommendation}} \\
\midrule
\multicolumn{2}{r|}{\# Users}&                    6,040 &                   25,741 &                     268 &                     346 \\
\multicolumn{2}{r|}{\# Items}&                    3,706 &                   23,643 &                   51,609 &                      20 \\
\multicolumn{2}{r|}{\# Interactions}&                 1,000,209 &                  708,919 &                  200,586 &                    1,879 \\
\multicolumn{2}{r|}{Min. User DEG}&                      20 &                       5 &                      21 &                       5 \\
\multicolumn{2}{r|}{Domain}&                   Movie &                 Grocery &                   Music &               Insurance \\
\cline{2-6}
\multirow{2}{*}{Repr.}& A &   O : 43.4\%; Y : 56.6\% &  O : 45.5\% ; Y : 54.5\% &   O : 42.2\%; Y : 57.8\% &   O : 49.4\%; Y : 50.6\% \\
& G & F : 28.3\%; M : 71.7\% &                     NA &   F : 42.2\%; M : 57.8\% &  F : 23.4\%; M : 76.6\% \\
\cline{2-6}
\multirow{2}{*}{$\overline{\text{User DEG}}$} & A &    O : 106.1; Y : 124.9 &      O : 20.7; Y : 19.6 &    O : 657.5; Y : 428.0 &        O : 4.3; Y : 4.5 \\
       & G &    F : 101.8; M : 122.7 &                  NA       &    F : 496.7; M : 545.3 &        F : 4.2; M : 4.5 \\
\cline{2-6}
\multirow{2}{*}{Gini User DEG} & A &      O : 0.53; Y : 0.52 &      O : 0.45; Y : 0.44 &      O : 0.43; Y : 0.43 &      O : 0.06; Y : 0.08 \\
       & G &      F : 0.53; M : 0.52 &                     NA    &      F : 0.42; M : 0.45 &      F : 0.05; M : 0.08 \\
\cline{2-6}
\multirow{2}{*}{$\overline{\text{User DY}}$} & A &      O : 0.07; Y : 0.07 &      O : 0.00; Y : 0.00 &      O : 0.03; Y : 0.03 &      O : 0.40; Y : 0.38 \\
       & G &      F : 0.07; M : 0.07 &                NA         &      F : 0.04; M : 0.04 &      F : 0.40; M : 0.39 \\
\cline{2-6}
\multirow{2}{*}{Gini User DY} & A &      O : 0.12; Y : 0.11 &      O : 0.45; Y : 0.42 &      O : 0.22; Y : 0.18 &      O : 0.12; Y : 0.14 \\
       & G &      F : 0.12; M : 0.12 &                    NA     &      F : 0.15; M : 0.23 &      F : 0.14; M : 0.13 \\
\cline{2-6}
\multirow{2}{*}{$\overline{\text{User IGD}}$} & A &      O : 0.95; Y : 0.96 &      O : 0.57; Y : 0.56 &      O : 0.99; Y : 0.99 &      O : 0.97; Y : 0.97 \\
       & G &      F : 0.95; M : 0.96 &                  NA       &      F : 0.99; M : 0.99 &      F : 0.97; M : 0.97 \\
\cline{2-6}
\multirow{2}{*}{Gini User IGD} & A &      O : 0.03; Y : 0.02 &      O : 0.05; Y : 0.05 &      O : 0.01; Y : 0.01 &      O : 0.01; Y : 0.01 \\
       & G &      F : 0.03; M : 0.02 &                  NA       &      F : 0.00; M : 0.01 &      F : 0.01; M : 0.01 \\

\bottomrule
\end{tabular}
}
\end{table}

Extensive research in user fairness in recommender systems is challenging due to the limited datasets including users' sensitive information.
We relied on the artifacts of a recent work accounting unfairness issues in recommendation \cite{10.1007/978-3-030-99736-6_37}, which performed a fairness assessment on two corpora: MovieLens 1M (ML-1M), on the movie domain, and Last.FM 1K (LFM-1K), on the music domain.
The time information of the users' interaction in LFM-1K was missing.
Hence, given that the interactions of each user are grouped by artist and considered as a single interaction, we set the timestamp of the last interaction with an artist to be the timestamp of the relative user-artist pair in LFM-1K.
We extended the set of datasets by including Insurance (INS), on the insurance domain, and Ta Feng (FENG), on the grocery domain\footnote{Yelp~\cite{DBLP:conf/recsys/MansouryMBP19} was also considered to include the business domain, but the users' gender information was predicted by their name, making questionable  analyses on this dataset.}.
All datasets include age and gender (except FENG) information for all users and their statistics are listed in Table \ref{tab:datasets}, where the graph properties values regard only the training set. The Gini coefficient for each property was measured as in \cite{DBLP:conf/icwsm/FabbriCB022}.

User nodes in INS and FENG were filtered by their number of interactions, i.e. their degree, so as to take into account users with histories made up of at least 5 items.
Duplicated interactions, e.g., users buying the same product twice in FENG, were removed.
On the basis of the binary setting mentioned in Section~\ref{subsec:lossoptim} and as done in \cite{10.1007/978-3-030-99736-6_37}, INS and FENG age labels were binarized as \emph{Younger (Y)} and \emph{Older (O)}, such that the \emph{Younger} group is more represented than the \emph{Older} one for consistency with ML-1M and LFM-1K.
Gender labels were already binary, as \emph{Males (M)} and \emph{Females (F)}.

We also adopted the splitting strategy used in \cite{10.1007/978-3-030-99736-6_37} for each dataset: per each user, 20\% (the most recent if a timestamp is available, randomly sampled otherwise) of the interactions forms the test set; the remaining interactions are split again, such that 10\% (selected in the same way) of this interactions subset forms the validation set, and the remaining 70\% forms the train set.
The validation set was used to select the training epoch where the model reported the best recommendation utility on the adjacency matrix $A$.
Given the goal of finding the edges causing unfairness in the test set, the truth ground labels of the latter were extracted to optimize the fairness loss in $\eqref{eq:fairloss}$.

\subsection{Models} \label{subsec:models}

% Which models did we use? Do they cover different types of GNNs?
Recently, novel GNNs have been devised to solve the top-$k$ recommendation task.
We relied on Recbole, which includes different families of GNN-based recommender systems.
\texttt{GNNUERS} was adopted on the following models:
\begin{itemize}
    \item GCMC \cite{DBLP:journals/corr/BergKW17}: this method is comprised of two components: a graph auto-encoder, which produces a node embedding matrix, and a decoder model, which predicts the relevance of the missing entries in the adjacency matrix from the node embedding matrix.
    \item NGCF \cite{DBLP:conf/sigir/Wang0WFC19}: this state-of-the-art GNN-based recommender system propagates embeddings in the user-item graph structure. In particular, it leverages high-order connectivities in the user-item integration graph, injecting the collaborative signal into the embeddings explicitly.
    \item LigthGCN \cite{DBLP:conf/sigir/0001DWLZ020}: it is a simplification of a GCN, including only the most essential components for collaborative filtering, i.e. the neighborhood aggregation. It uses a single embedding as the weighted sum of the user and item embeddings propagated at all layers in the graph.
\end{itemize}
The GNNs were trained with the default and model-specific hyper-parameters defined by Recbole.

\subsection{Explanation Baseline Methods} \label{subsec:baselines}

As mentioned in Section \ref{subsec:graphgen}, \texttt{GNNUERS} in its base form applies a policy that prevents the algorithm from restoring previously deleted edges.
Additionally, we examined an extension of \texttt{GNNUERS} by applying another policy, \emph{Connected Nodes (CN)}. It limits the perturbation to the edges connected to the unprotected user nodes to investigate whether recommendation unfairness is only due to the interactions performed by the unprotected group.
Therefore, \texttt{GNNUERS+CN} guides the learning process to select the users' actions of the unprotected group that made $f$ favor them.

The literature does not include baselines that explain unfairness in the form of user-item edges as \texttt{GNNUERS}.
The works proposing unfairness explainability methods in recommendation \cite{DBLP:journals/ipm/DeldjooBN21,DBLP:conf/sigir/GeTZXL0FGLZ22} select relevant user/item features as explanations, which cannot be compared with the ones generated by our framework.
Approaches proposed to explain unfairness in GNNs \cite{DBLP:conf/aaai/DongW0LL23} were devised for classification tasks.
Although it is not clear if such methods could be extended to recommendation tasks, such an engineering adaptation goes beyond the goal of our work.
{\color{black} The alternative counterfactual explainability algorithms in GNNs present limitations as well.
Specifically, methods that generate explanations at the model level~\cite{DBLP:conf/wsdm/HuangKMRS23} present computational and memory footprint limitations that prevent their extension from small graphs, e.g., molecules, to the typically larger networks of user-item interactions.
The methods that generate explanations at the instance level~\cite{DBLP:conf/aistats/LucicHTRS22,DBLP:conf/pkdd/KangLB21} cannot be trivially} adapted to envision the unfairness task at the model level.

To this end, we adopted \texttt{CASPER} \cite{DBLP:conf/cikm/OhUMK22} for comparison, a model-agnostic method that causes the highest instability in the recommendations by perturbing a single interaction, i.e. an edge of the graph.
The instability induced by \texttt{CASPER} could alter the recommendations, and, as a result, re-distributing the utility levels over the demographic groups and positively affecting unfairness.
At inference step, our models generate the recommendation lists by using the training network perturbed by \texttt{CASPER}, then fairness and utility metrics are measured.
\texttt{CASPER} uses the timestamp of each interaction to generate a directed acyclic graph of the interactions of each user.
INS does not include the time information, so \texttt{CASPER} was not applied on this dataset.

We also introduce \texttt{RND-P} as sanity check, a baseline algorithm that at each iteration randomly perturbs edges with a probability $\rho$, such that it mimics the \texttt{GNNUERS} edges selection process, but based on a random choice.
Given the size diversity of our evaluation datasets, we set $\rho = 1 / (|E_{train}| / 100) $, where $E_{train}$ is the set of training edges, as the value that works best across the selected epochs.
As a consequence, \texttt{RND-P} perturbs edges depending on the network size to prevent the random sampling from deleting all the edges in a few iterations.

The explanation methods were executed on all the models and datasets over 800 epochs adopting an early stopping method when $\mathcal{L}_{fair}$ does not improve with a delta higher than 0.001 for at least 15 consecutive epochs.
The best hyperparameters (learning rate, $\beta$, batch size) can be found in the source code repository.
{\color{black} In order to preserve the recommendation scenario described by the original adjacency matrix, \texttt{GNNUERS} stops if all the interactions of a user or item are deleted.
Nonetheless, this constraint was never triggered in our experiments.}

{\color{black} The following experiments employ the selected datasets and models to assess the ability of \texttt{GNNUERS} of explaining unfairness in GNN-based recommendation. In view of this goal, we expect our method to be used in scenarios where models exhibit a certain degree of unfairness, which is determined by the context and the requirements imposed by platform owners and policy makers. Therefore, in our experiments, we assume the recommendations generated by the selected models to be unfair to some extent, even subtly, on the target datasets.}

\subsection{RQ1: Unfairness Explainability Benchmark} \label{subsec:unfair_exp_bench}

\begin{figure}[!b]
    \resizebox{\linewidth}{!}{\begin{tabular}{cc}
        \multicolumn{2}{c}{\includegraphics[width=\linewidth]{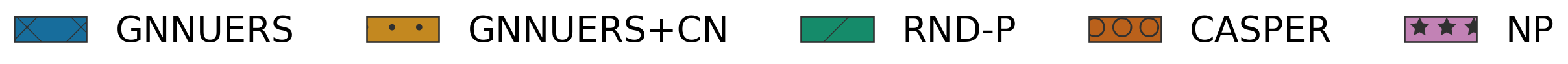}} \\
        \includegraphics[width=\linewidth]{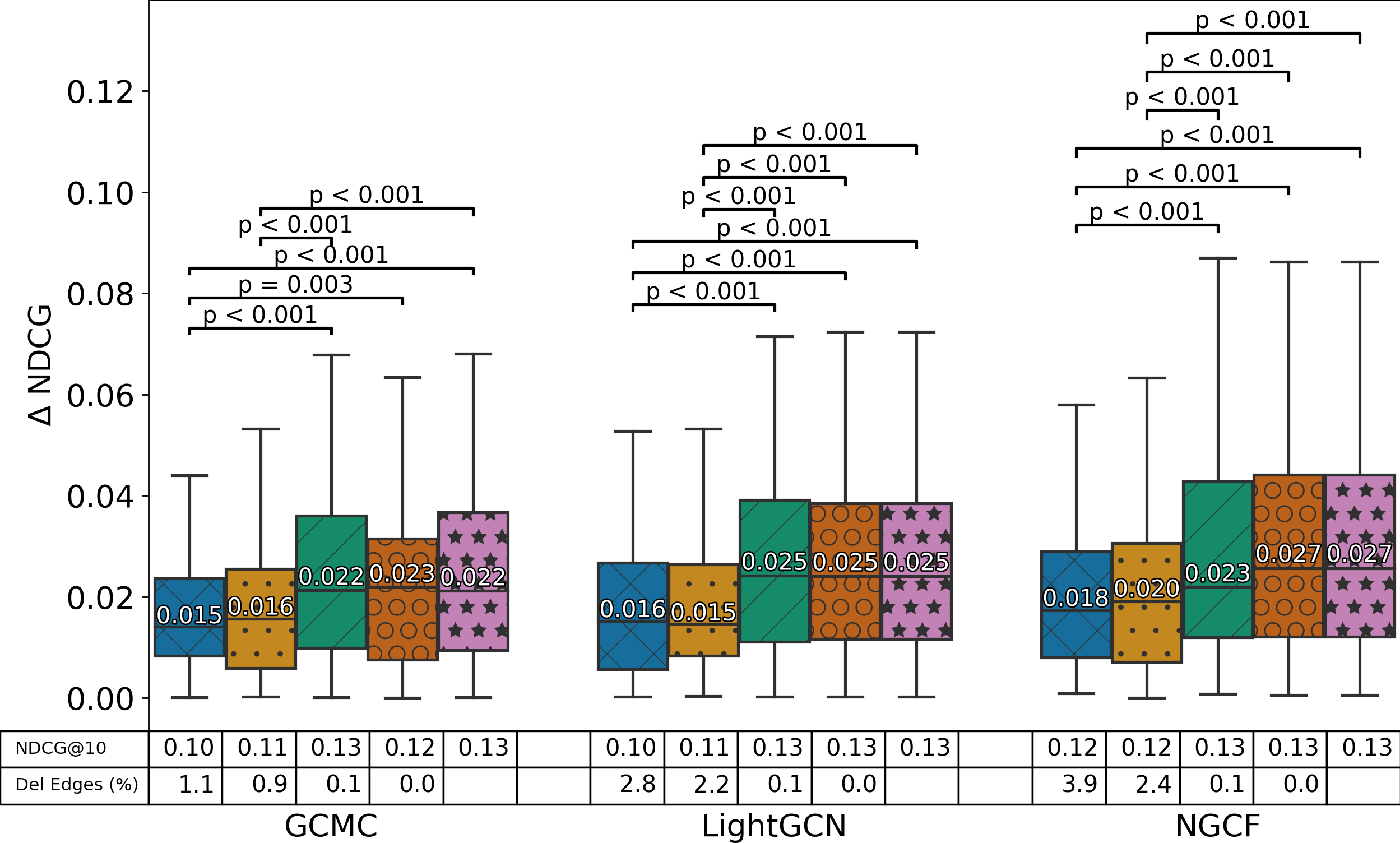} & \includegraphics[width=\linewidth]{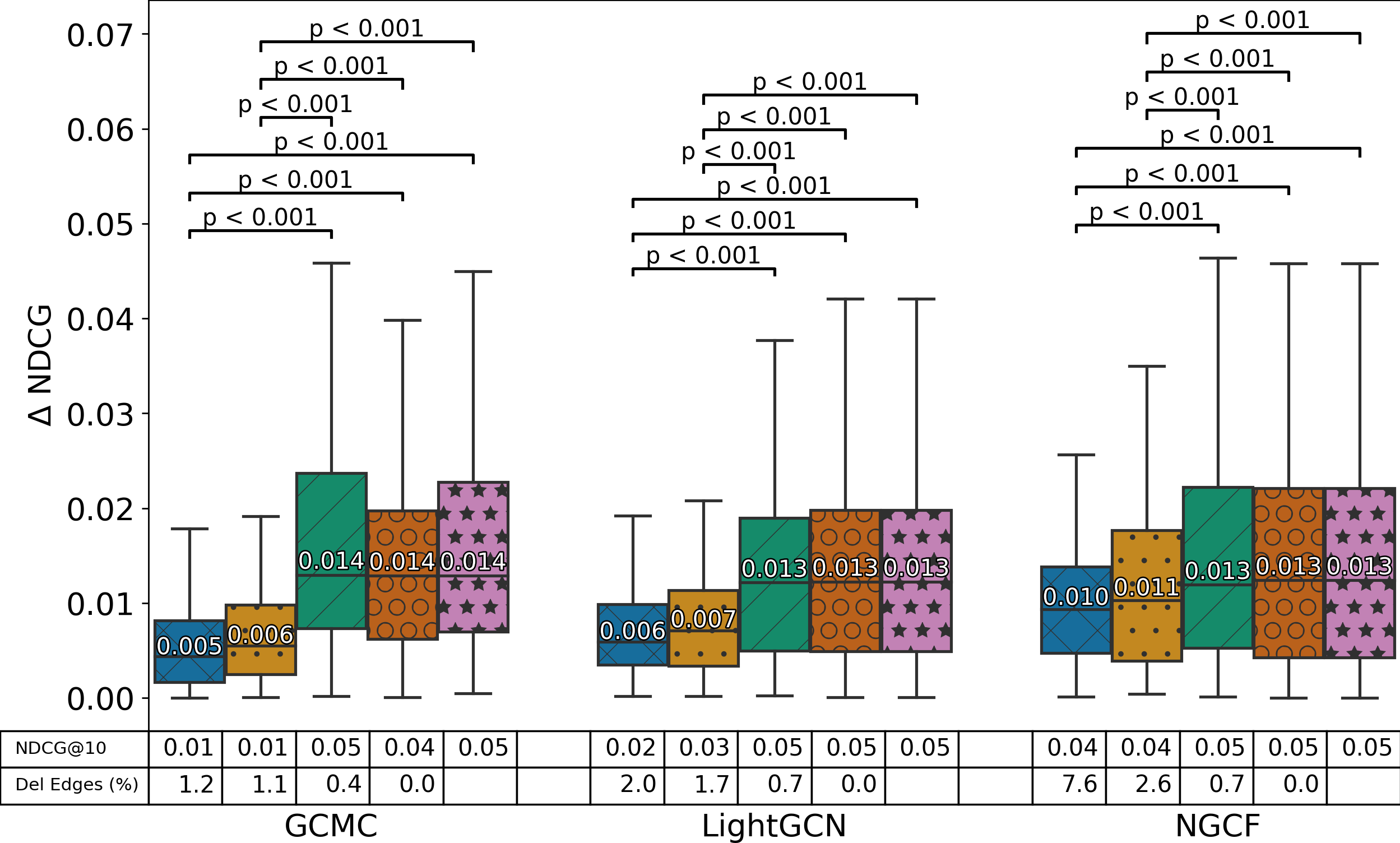} \\
        ML-1M & FENG\\
        \includegraphics[width=\linewidth]{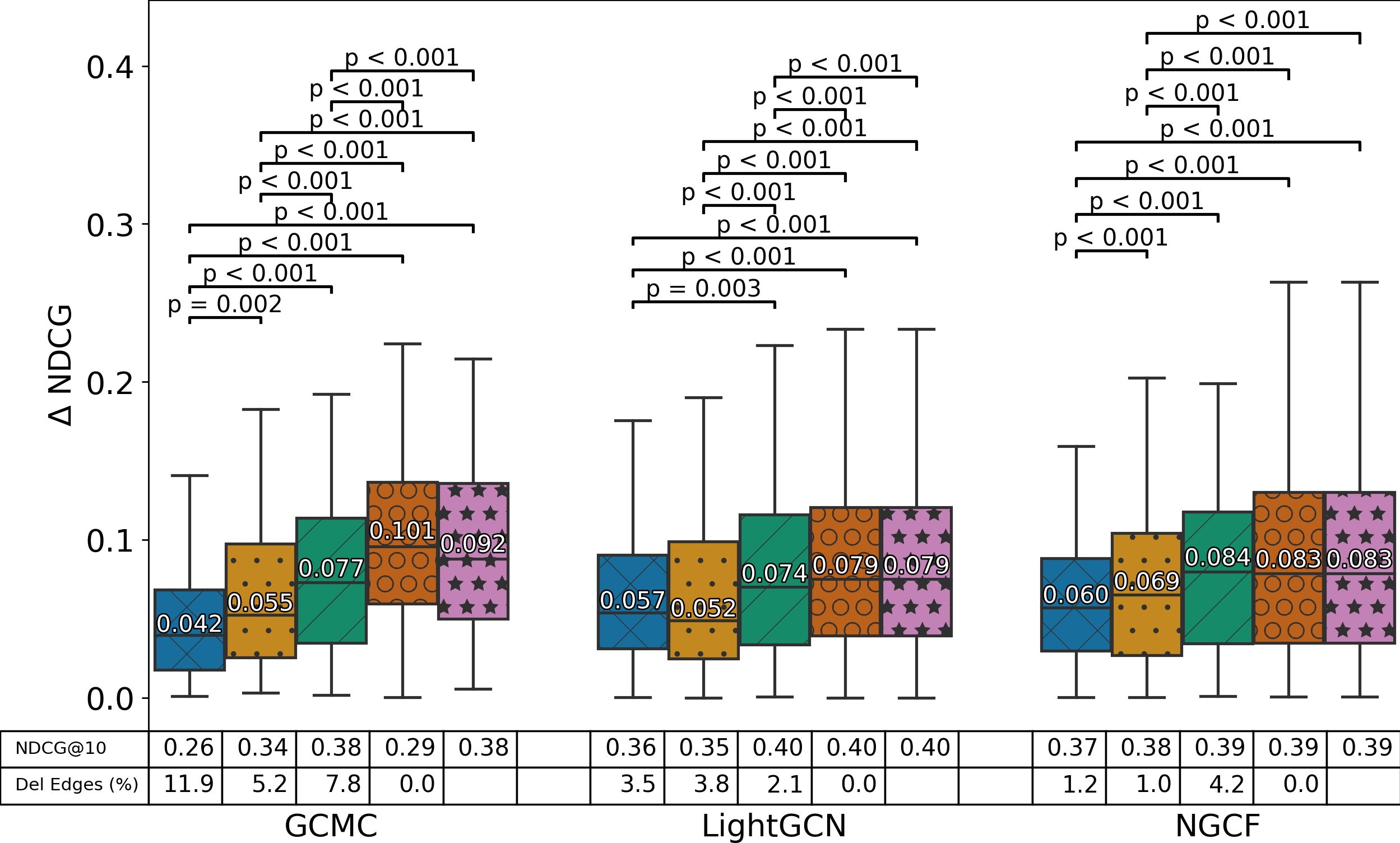} & \includegraphics[width=\linewidth]{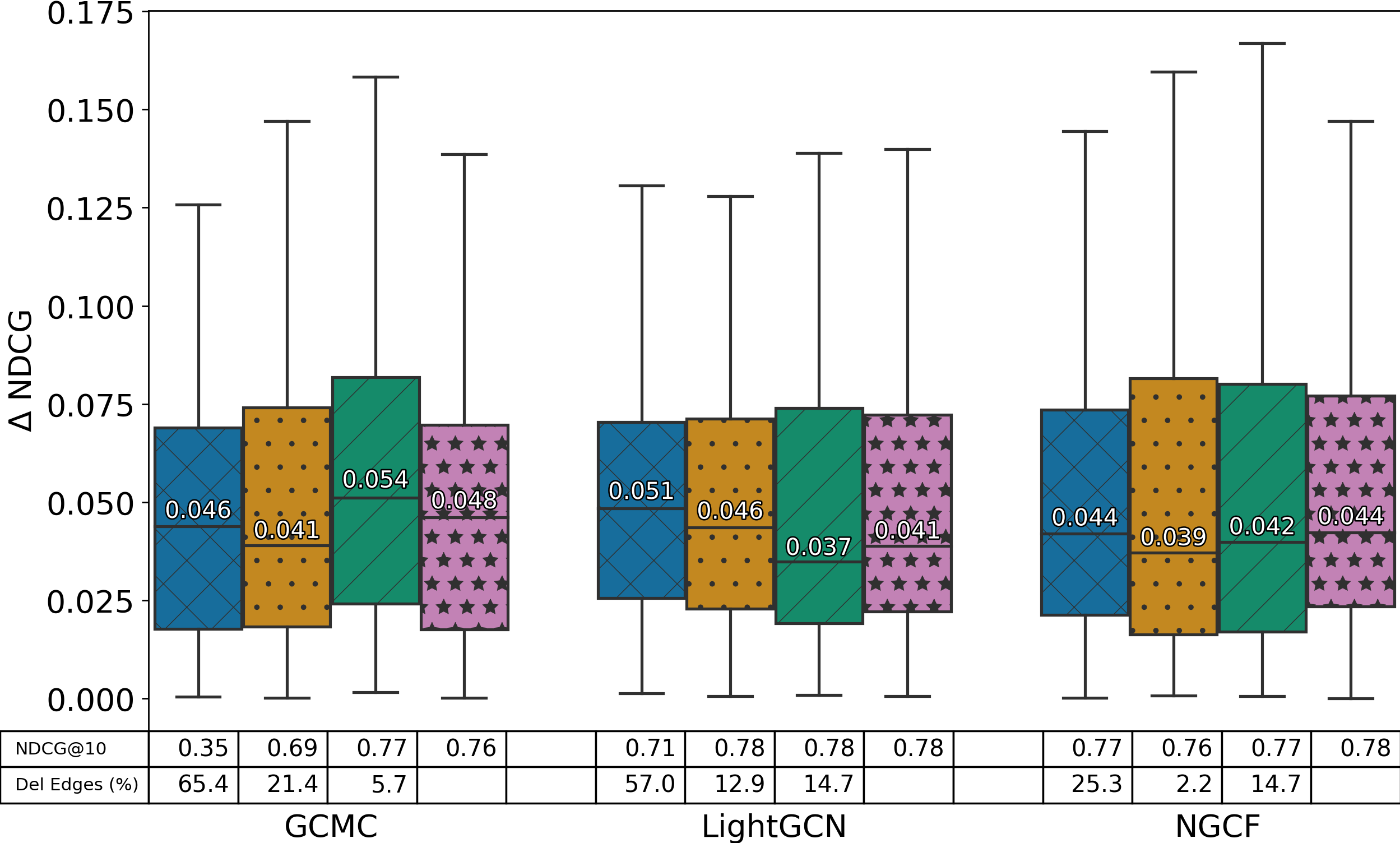} \\
        LFM-1K & INS \\
    \end{tabular}}
    \caption{Distribution of $\Delta$NDCG between younger and older users subgroups, randomly sampled 100 times. A Wilcoxon signed-rank test is performed between each pair of boxes and the respective p-value is shown if it is lower than $\frac{0.05}{m}$ according to the Bonferroni correction, where $m$ is the number of pairwise comparisons performed for each model.}
    \Description{}
    \label{fig:deltandcg_age}
\end{figure}

\begin{figure*}[!b]
    \resizebox{\linewidth}{!}{\begin{tabular}{cc}
        \multicolumn{2}{c}{\includegraphics[width=\linewidth]{images/Legend.png}} \\
        \includegraphics[width=\linewidth]{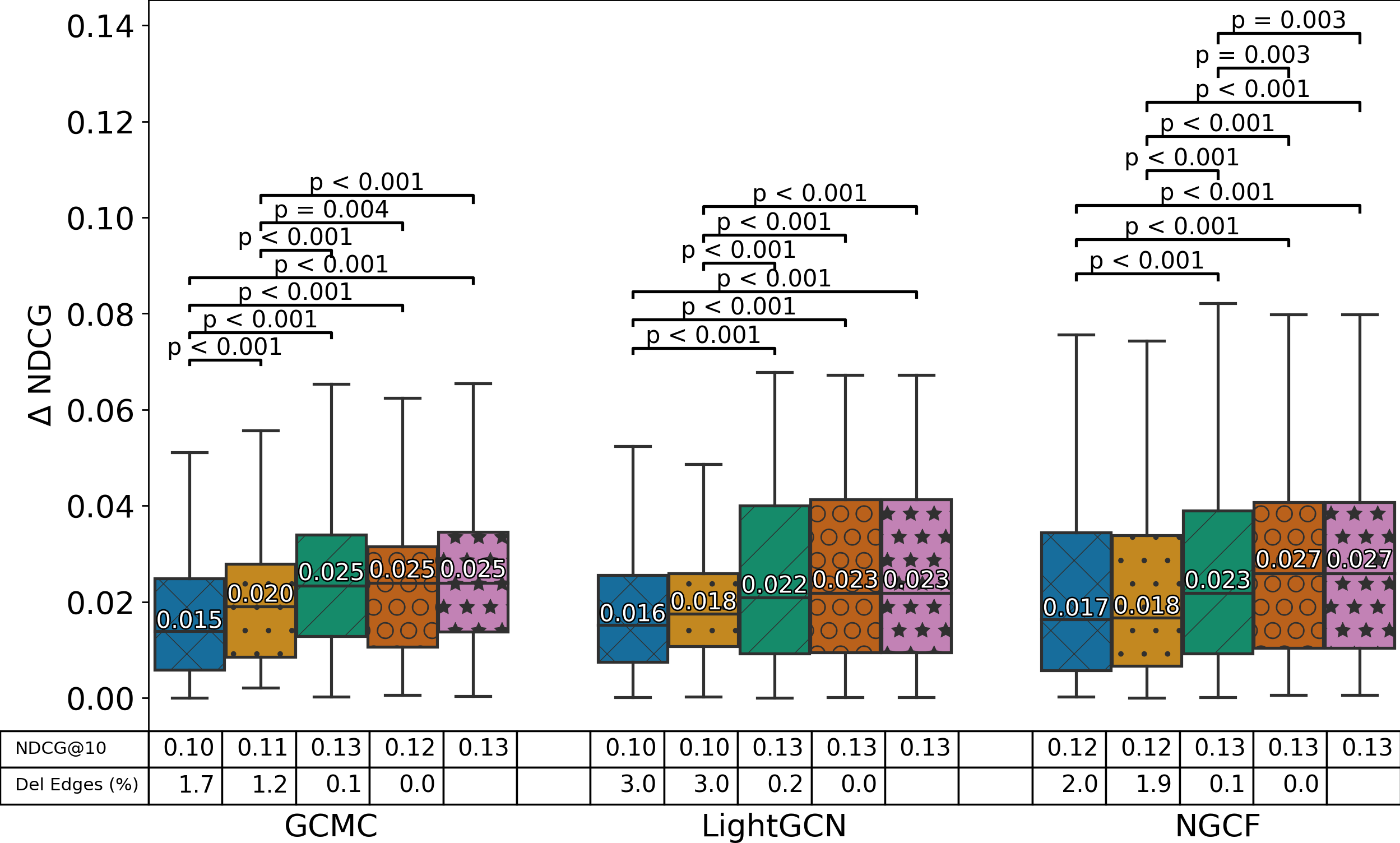} & \includegraphics[width=\linewidth]{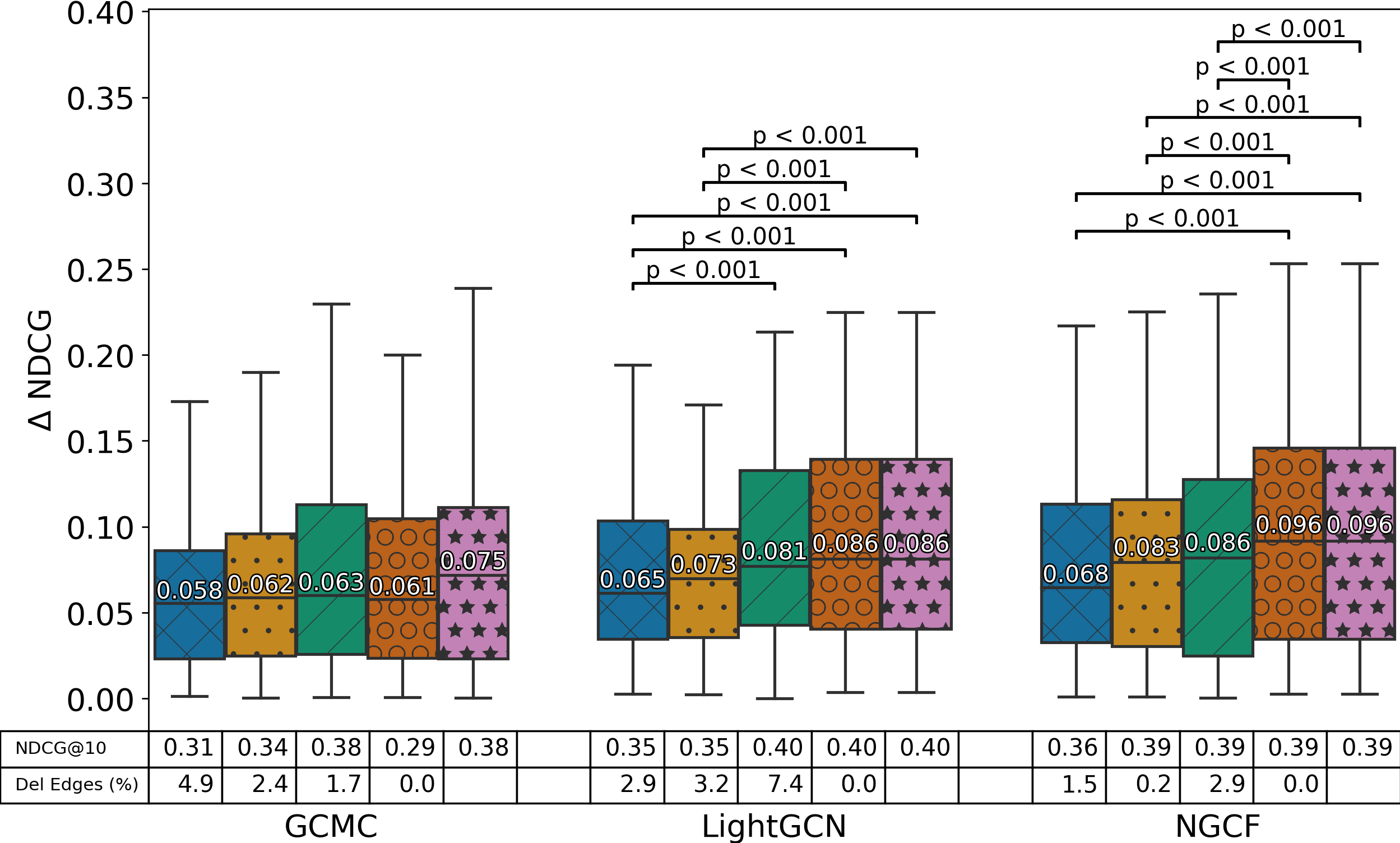} \\
        ML-1M & LFM-1K \\
        \multicolumn{2}{c}{\includegraphics[width=\linewidth]{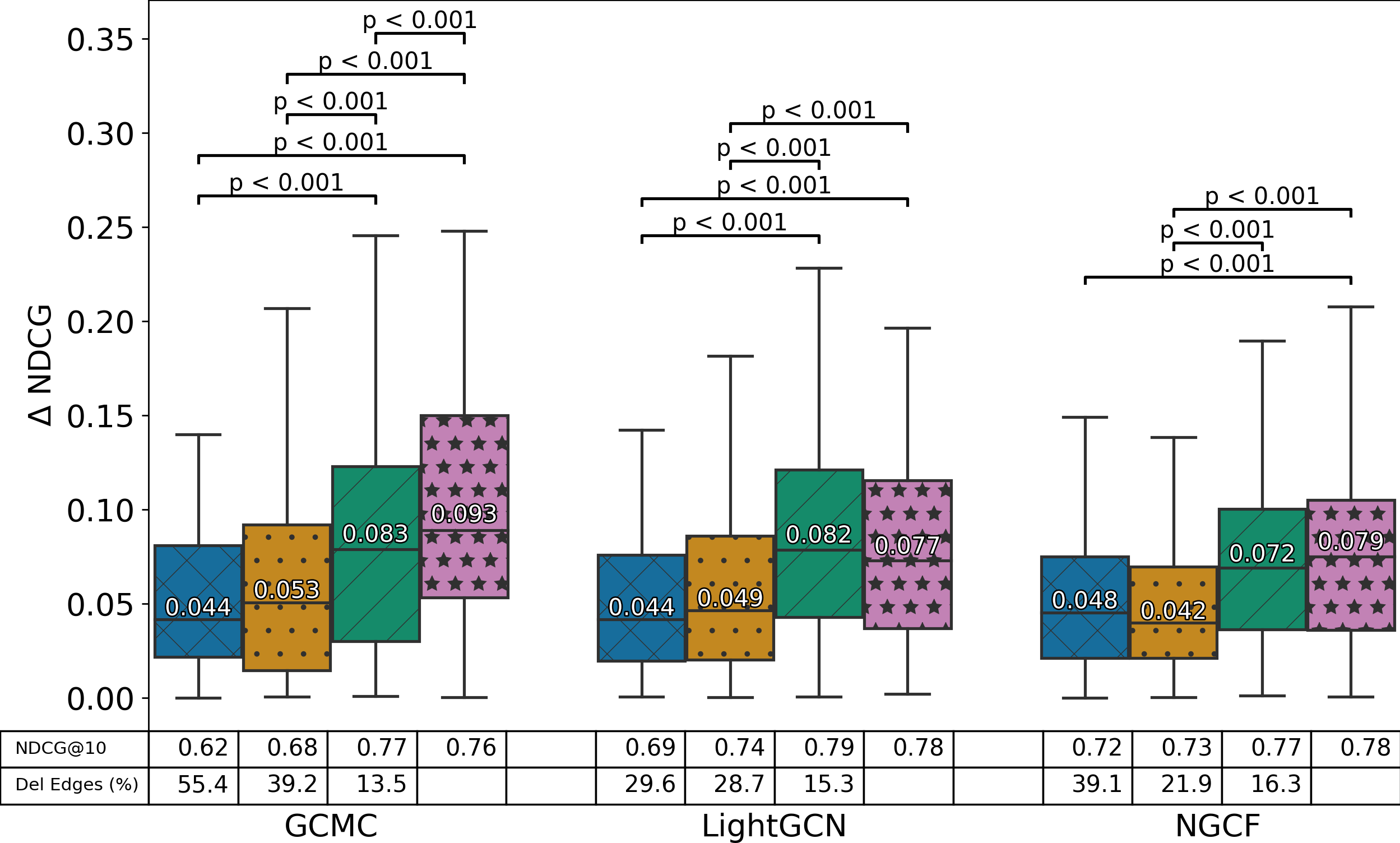}} \\
        \multicolumn{2}{c}{INS}\\
    
    \end{tabular}}
    \caption{Distribution of $\Delta$NDCG between males and females users subgroups, randomly sampled 100 times. A Wilcoxon signed-rank test is performed between each pair of boxes and the respective p-value is shown if it is lower than $\frac{0.05}{m}$ according to the Bonferroni correction, where $m$ is the number of pairwise comparisons performed for each model.}
    \Description{}
    \label{fig:deltandcg_gender}
\end{figure*}

% {RQ1}: Can the perturbation of the graph topological structure mitigate recommendation utility unfairness under the operationalised fairness notion
We first investigated the capability of \texttt{GNNUERS} to select counterfactual explanations that effectively optimize \eqref{eq:fairloss}.
The \texttt{GNNUERS} learning process selects users coming from both demographic groups, stores them in fixed-size batches according to their distribution in the dataset, and optimizes the loss to minimize disparity in NDCG@10 (average) between the protected and unprotected group.
The evaluation phase follows an analogous process: we randomly sample 100 subgroups with the same demographic groups distribution of the batches and with sample size equal to the batch size.
This choice is also due to reduce the sampling bias present in the datasets, i.e. the evaluation is not affected by the different sample size of unprotected and protected groups.
The batch size for each dataset is selected such that it splits the users in at least five partitions, guaranteeing a low probability of picking the same users in the randomly sampled subgroups.
We measure the differences in each subgroup with $\Delta$NDCG, which corresponds to the difference in performance between the two user groups.
$\Delta$NDCG is then related to the unfairness level measured by \eqref{eq:fairloss} (without the modulus operator){\color{black}, and the function $\Phi$ used in Definition \ref{def:cf_exp_unfair}-\ref{def:opt_cf_exp_unfair}}.
%If the edges perturbation reduces $\Delta$NDCG across all the subgroups, such edges represent the cause of the utility disparity between demographic groups. 
%
We compare our proposed solution with \texttt{CASPER}, \texttt{RND-P} and the original $\Delta$NDCG values, {\color{black} denoted as \texttt{NP}}.

\noindent \textbf{Age.} Figure \ref{fig:deltandcg_age} shows the $\Delta$NDCG distribution across the subgroups defined by the attribute "age".
We adopt the Wilcoxon signed-rank test and report, on top of each boxplot, the p-value significance of the difference between the means of each distributions pair (it is reported IFF the p-values are lower than $\frac{0.05}{m}$ according to the Bonferroni correction, where $m$ is the number of pairwise comparisons performed for each model). 
On the bottom of the plots, we include the NDCG@10 after the perturbation and the percentage of deleted edges.
First, we can see how in ML-1M and FENG our methods significantly narrow the $\Delta$NDCG distribution of the age subgroups compared with \texttt{NP}, by perturbing just 1\% of the edges in some settings.
The single edge deleted by \texttt{CASPER} minimally affects the recommendations of all the experiments, with more prominent effects on GCMC under several datasets, e.g., ML-1M, FENG.
On the other hand, the perturbation applied by \texttt{RND-P}, in some cases, can generate a decrease in unfairness, but the early stopping prevents it from removing too many edges, which highlights its inefficiency in reducing $\Delta$NDCG along the epochs.
Among our methods, \texttt{GNNUERS+CN} reports the least perturbations and generates competitive decreases in $\Delta$NDCG.
On INS, no explanation method is consistent in reducing $\Delta$NDCG.
This means that the difference in performance between the groups derived from "age" are not affected by the perturbation of individual interactions, but other aspects (e.g., specific nodes) possibly caused the original disparity.

Generally, \texttt{GNNUERS} and \texttt{GNNUERS+CN} are able to significantly reduce $\Delta$NDCG in most of the settings, selecting small subsets of deleted edges.
Interestingly, in some cases, the NDCG@10 drops significantly (LFM-1K and FENG), while in others remains consistent (ML-1M). 
This means that our algorithms are able to generate explanations of the unfairness by detecting edges that contribute significantly to increasing unfairness and to improving performance for only one subgroup.
%Even removing more than 60\% of edges is not enough to mitigate unfairness by rearranging 20 insurance products in 10-item recommendation lists.

\noindent \textbf{Gender.} \texttt{GNNUERS} generates a significant decrease in $\Delta$NDCG also for the subgroups generated by the attribute "gender", as shown in Figure \ref{fig:deltandcg_gender}.
In ML-1M and LFM-1K, \texttt{GNNUERS} significantly explains unfairness for all the models, reducing $\Delta$NDCG by a relatively lower amount compared to the same experiments on age groups (e.g., on LFM-1K).
%modifying the network topologies by an even lower number of edges compared to the same experiments on age groups.
While our method has proven to be effective regardless of the sensitive attribute, perturbing a single edge (\texttt{CASPER}) or more in a random way (\texttt{RND-P}) does not decrease $\Delta$NDCG in these cases (except for GCMC on LFM-1K).
On INS, differently from what seen before, \texttt{GNNUERS} can reduce unfairness between gender groups, by deleting a relevant higher number of edges compared to other experiments.
This result emphasizes how the users' splitting in groups could result in different types of unfairness.
%The more biased representation of gender groups (Table \ref{tab:datasets}) permits \texttt{GNNUERS} to better distinguish the relevant edges for the unprotected group, but there is still no consistency across models.

\hlbox{RQ1}{
Except extreme cases, \texttt{GNNUERS} selects edges that systematically and significantly explain unfairness, regardless of the data, models and user groups on which is applied.
}

As shown in both demographic groups, NDCG is not always drastically impacted globally.
For this reason, we run a more thorough analysis of the impact on utility.

\subsection{RQ2: Impact on Recommendation Utility}

\begin{table}[!b]
    \caption{For both protected and unprotected groups, each column includes the value of NDCG after applying \texttt{GNNUERS} and in the brackets its relative change from the original NDCG. Unproctected group values are highlighted and in bold.}
    \resizebox{.75\linewidth}{!}
    {\begin{tabular}{lll|rr|rr}
    \toprule
    \midrule
           & & & \multicolumn{2}{c|}{Age} & \multicolumn{2}{c}{Gender} \\
            & Model & Policy &                   \multicolumn{1}{c}{Younger}  &                    \multicolumn{1}{c|}{Older}                    &      \multicolumn{1}{c}{Males}      &        \multicolumn{1}{c}{Females}  \\
    \midrule
    \multirow{6}{*}{\rotatebox[origin=c]{90}{ML-1M}} & \multirow{2}{*}{GCMC} & \texttt{GNNUERS} &  \textBF{\hl{0.11* (-21.6\%)}} &             0.11* (-11.6\%)  &                        \textBF{\hl{0.10* (-24.2\%)}} &                        0.10* (-14.6\%)  \\
          &      & \texttt{GNNUERS+CN} &             \textBF{\hl{0.12* (-14.0\%)}} &             0.11* (-06.7\%)  &                        \textBF{\hl{0.11* (-16.3\%)}} &                        0.10* (-10.2\%)  \\
        \cline{2-7}
          & \multirow{2}{*}{LightGCN} & \texttt{GNNUERS} &             \textBF{\hl{0.11* (-21.6\%)}} &             0.10* (-15.4\%)  &                        \textBF{\hl{0.10* (-22.8\%)}} &                        0.09* (-17.6\%) \\
          &      & \texttt{GNNUERS+CN} &             \textBF{\hl{0.11* (-16.7\%)}} &             0.10* (-10.6\%)  &                        \textBF{\hl{0.11* (-19.4\%)}} &                        0.09* (-14.1\%)  \\
        \cline{2-7}
          & \multirow{2}{*}{NGCF} & \texttt{GNNUERS} &             \textBF{\hl{0.12* (-11.6\%)}} &             0.11* (-02.7\%)  &                        \textBF{\hl{0.13* (-07.8\%)}} &             0.11* (-01.8\%) \\
          &      & \texttt{GNNUERS+CN} &             \textBF{\hl{0.13* (-09.2\%)}} &  0.12\hspace{3.5pt} (-00.4\%)  &                        \textBF{\hl{0.13* (-07.1\%)}} &             0.11\hspace{3.5pt} (-01.0\%) \\
        \cline{1-7}
        \cline{2-7}
        \midrule
        \multirow{6}{*}{\rotatebox[origin=c]{90}{FENG}} & \multirow{2}{*}{GCMC} & \texttt{GNNUERS} &             0.01* (-67.0\%) &             \textBF{\hl{0.01* (-78.9\%)}}  &                                     - &                                     - \\
          &      & \texttt{GNNUERS+CN} &             0.02* (-62.4\%) &             \textBF{\hl{0.01* (-77.6\%)}}  &                                     - &                                     -  \\
        \cline{2-7}
          & \multirow{2}{*}{LightGCN} & \texttt{GNNUERS} &             0.02* (-51.3\%) &             \textBF{\hl{0.02* (-61.2\%)}}  &                                     - &                                     -  \\
          &      & \texttt{GNNUERS+CN} &             0.03* (-36.7\%) &             \textBF{\hl{0.03* (-51.2\%)}}  &                                     - &                                     -  \\
        \cline{2-7}
          & \multirow{2}{*}{NGCF} & \texttt{GNNUERS} &             0.04* (-10.4\%) &             \textBF{\hl{0.04* (-28.1\%)}}  &                                     - &                                     -  \\
          &      & \texttt{GNNUERS+CN} &             0.04* (-01.9\%) &             \textBF{\hl{0.05* (-13.2\%)}}  &                                     - &                                     - \\
        \cline{1-7}
        \cline{2-7}
        \midrule
        \multirow{6}{*}{\rotatebox[origin=c]{90}{LFM-1K}} & \multirow{2}{*}{GCMC} & \texttt{GNNUERS} &             0.27* (-22.8\%) &             \textBF{\hl{0.26* (-40.4\%)}}  &                        0.33* (-12.0\%) &                        \textBF{\hl{0.31* (-26.4\%)}} \\
          &      & \texttt{GNNUERS+CN} &             0.32* (-08.0\%) &             \textBF{\hl{0.36* (-16.4\%)}}  &                        0.35* (-05.9\%) &                        \textBF{\hl{0.35* (-17.4\%)}}  \\
        \cline{2-7}
          & \multirow{2}{*}{LightGCN} & \texttt{GNNUERS} &             0.34* (-05.9\%) &             \textBF{\hl{0.37* (-15.5\%)}}  &                        0.34* (-07.9\%) &                        \textBF{\hl{0.37* (-16.7\%)}} \\
          &      & \texttt{GNNUERS+CN} &             0.33* (-09.5\%) &             \textBF{\hl{0.37* (-15.8\%)}}  &                        0.35* (-07.4\%) &                        \textBF{\hl{0.38* (-13.9\%)}}  \\
        \cline{2-7}
          & \multirow{2}{*}{NGCF} & \texttt{GNNUERS} &  0.35* (-01.5\%) &             \textBF{\hl{0.39* (-09.6\%)}} &  0.36\hspace{3.5pt} (-00.9\%) &                        \textBF{\hl{0.38* (-12.4\%)}} \\
          &      & \texttt{GNNUERS+CN} &  0.35\hspace{3.5pt} (\hspace{3pt}00.2\%) &             \textBF{\hl{0.40* (-05.9\%)}}  &                        0.37* (\hspace{3pt}02.5\%) &                        \textBF{\hl{0.42* (-04.1\%)}} \\
        \cline{1-7}
        \cline{2-7}
        \midrule
        \multirow{6}{*}{\rotatebox[origin=c]{90}{INS}} & \multirow{2}{*}{GCMC} & \texttt{GNNUERS} &             \textBF{\hl{0.37* (-52.7\%)}} &             0.33* (-56.8\%)  &                        \textBF{\hl{0.62* (-20.0\%)}} &                        0.60* (-13.3\%) \\
          &      & \texttt{GNNUERS+CN} &             \textBF{\hl{0.69* (-10.6\%)}} &    0.70* (-08.0\%)  &                        \textBF{\hl{0.68* (-12.4\%)}} &                        0.67* (-02.7\%)  \\
        \cline{2-7}
          & \multirow{2}{*}{LightGCN} & \texttt{GNNUERS} &             \textBF{\hl{0.71* (-09.9\%)}} &             0.71* (-08.4\%) &                        \textBF{\hl{0.69* (-12.8\%)}} &                        0.69* (-05.2\%)  \\
          &      & \texttt{GNNUERS+CN} &             \textBF{\hl{0.79 \, ( 00.1\%)}} &  0.78\hspace{3.5pt} (\hspace{3pt}00.3\%)  &                        \textBF{\hl{0.74* (-06.7\%)}} &                        0.73* (\hspace{3pt}01.5\%)  \\
        \cline{2-7}
          & \multirow{2}{*}{NGCF} & \texttt{GNNUERS} &             0.78\hspace{3pt} (-00.3\%) &             \textBF{\hl{0.77* (-02.4\%)}}  &      \textBF{\hl{0.72* (-09.6\%)}} &             0.73* (-01.2\%) \\
          &      & \texttt{GNNUERS+CN} &  0.77* (-01.1\%) &  \textBF{\hl{0.76* (-03.8\%)}}  &                        \textBF{\hl{0.73* (-08.7\%)}} &  0.74\hspace{3.5pt} (\hspace{3pt}00.7\%)  \\
    \midrule
    \bottomrule
    \end{tabular}
    }
    \label{tab:ndcg_before_after}
\end{table}

% \item \textbf{RQ2}: Does the edges deletion reduce the unprotected group utility while minimally affecting the protected group one?

{\color{black} Our unfairness explanation task aims to identify the user-item interactions that made the recommender system advantage a demographic group, i.e. the unprotected group.
Given that the same user-item interactions represent the main source of information for recommender systems, we expect the edges deletion process of \texttt{GNNUERS} to negatively affect the recommendation utility to find a proper explanation.
To this end,} \texttt{GNNUERS} is devised to minimize the gap in recommendation utility between the demographic groups, but without or minimally affecting the utility for the protected group.
{\color{black} This perspective is related to Definition~\ref{def:opt_cf_exp_unfair}, such that a faithful explanation would properly uncover the interactions that advantaged the unprotected group}.
We empirically evaluate this aspect, by examining the edges deletion impact on the utility for each group.
The NDCG@10 was measured individually for each user to then averaging it across demographic groups.
The impact on recommendation utility was measured as the change in utility after applying the perturbation.
To estimate this change significance, we first leveraged the 100 subgroups used in Section~\ref{subsec:unfair_exp_bench}.
Then, for each subgroup, we gathered the NDCG@10 average measured on the recommendations altered from each explanation method and the average generated from the non-perturbed network.
Finally, a Wilcoxon signed-rank test was performed between the averages resulting from the non-perturbed network and from each explanation method.
For this analysis, we consider \texttt{GNNUERS} and \texttt{GNNUERS+CN}.

Table~\ref{tab:ndcg_before_after} shows the average utility (highlighted for the unprotected group) after perturbing the edges and its relative change from the original one between brackets.
For each value, the symbol (*) denotes the significance of the statistical test with the 95\% of confidence interval.
We can see how, for any dataset and model, the NDCG change for the unprotected group is greater than the protected group one.
This confirms that our algorithms can select the edges responsible for an higher utility for the unprotected group.
However, also the NDCG for the protected group is affected in most of the experiments. This is because the results are model dependent, and removing edges reduces the connectivity, and then the information propagation through the GNN.
Also, higher NDCG losses for the unprotected groups reflect a better unfairness explanation, as seen for all the models in FENG\footnote{\texttt{GNNUERS} learning process could be stopped once a desired level of fairness or utility is reached for the explanation, depending on the application requirements.}.
Based on this observation, since \texttt{GNNUERS} perturbations for NGCF result in the least faithful unfairness explanation w.r.t. the other models in the previous RQ, for the same model it reports the lowest loss in utility for both demographic groups.
As a matter of fact, not only for NGCF the NDCG for the protected group is minimally affected, but it also increases in some settings, e.g., for males users in LFM-1K.
The \texttt{GNNUERS+CN} policy enforces this behavior more than \texttt{GNNUERS} by a slight amount, except for FENG.
Indeed, in FENG the NDCG is equal between demographic groups, but \texttt{GNNUERS} reports an additional 10\% loss in utility.
Using \texttt{GNNUERS+CN} edges selection is then beneficial to reduce the impact on the NDCG for the protected group.

\hlbox{RQ2}{
\texttt{GNNUERS} and \texttt{GNNUERS+CN} reduce the utility for unprotected groups, detecting edges that led to a disparity in performance, while reporting a negligible loss for the protected ones.
}

\subsection{RQ3: Edges Selection Process}  \label{subsec:edge_selec_process}

The findings of the first two research questions regard the \texttt{GNNUERS} performance in explaining the unfairness and its impact on the recommendation utility of each demographic group.
Despite the edges selected by \texttt{GNNUERS} represent themselves a counterfactual explanation of the utility disparity across demographic groups, the edges taken as they are do not provide sufficient information to fully understand the graph features that contribute to the unfairness.
To this purpose, we leverage the user nodes connected to such edges and categorize them by the properties defined in Section \ref{subsec:graph_properties}.
Specifically, user nodes were distinguished by demographic groups (\emph{Males}, \emph{Females}, \emph{Younger}, \emph{Older}) and the nodes in each group were partitioned in quartiles.
The order of the data points was defined by the value of each graph property that characterizes the users in each quartile, e.g., low-DEG nodes (Q1).
Thus, the graph properties were individually measured for each demographic group by defining the subgroups \emph{Males}, \emph{Females}, \emph{Younger}, \emph{Older} as the set $Z$ one at a time.
For instance, if $Z$ is the subgroup of \emph{Males}, $IGD_z$ represents how the male $z \in Z$ is close to the other males $z' \in Z/\{z\}$.

For each demographic group and each quartile, the number of perturbed edges was taken and normalized by the total number of edges perturbed in each experiment.
Therefore, the perturbed edges can be categorized as well based on the extent such edges are distributed over the quartiles for each graph property.
A higher distribution over nodes with low or high DEG, DY, IGD levels would highlight the aspects that drove our explanation method to perturb the given edges, and, therefore, highlight the aspects that led to the unfairness in the original recommendations.

\texttt{GNNUERS+CN} was not included in the following experiments because the edges deletion constraint could only reflect the distribution over quartiles defined for the unprotected group.
RQ1 reported FENG, ML-1M, LFM-1K as the datasets on which our method consistently explain unfairness by reducing the disparity in NDCG between the demographic groups derived by both "gender" and "age".
RQ2 highlighted NGCF as the model where \texttt{GNNUERS} can explain unfairness while affecting the recommendation utility of the protected group the least.
\texttt{GNNUERS} selection process will be analyzed only on the experiments with these settings, as they would better reflect the properties characterizing biased recommendations.

\begin{figure*}[!b]
    \includegraphics[width=\linewidth]{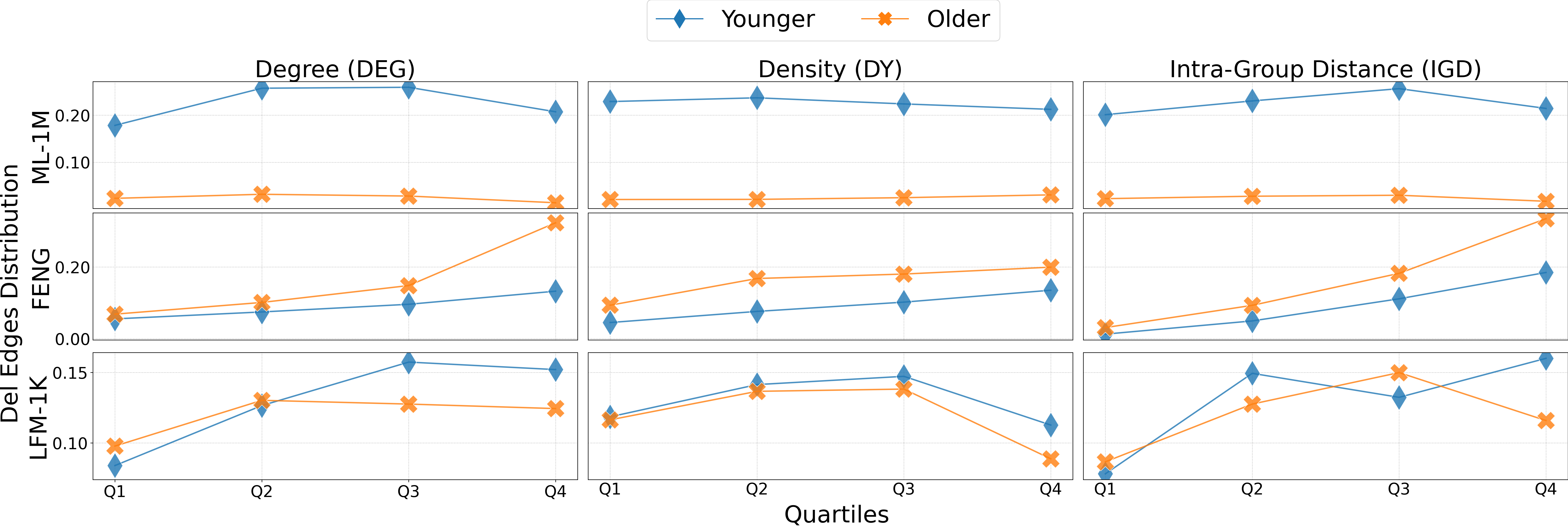}
    \caption{Deleted edges distribution (\emph{Del Edges Distribution}) over the quartiles (Q1-Q2-Q3-Q4) defined for each \emph{age} group through sorting the nodes by each graph property. The edges were deleted applying \texttt{GNNUERS} on NGCF.}
    \Description{}
    \label{fig:graph_metric_ngcf_age}
\end{figure*}

\noindent \textbf{Age.} For each graph property, Figure~\ref{fig:graph_metric_ngcf_age} reports the distribution of the edges deleted by \texttt{GNNUERS} over the quartiles, which are defined for each demographic group derived from "age".
Except for LFM-1K, the distribution of deleted edges is significantly higher for the unprotected groups (ML-1M: Younger, FENG: Older), which highlights that the unfair recommendations are highly dependent on the unprotected users' interactions.
On ML-1M, the over-representation (Table~\ref{tab:datasets}) and over-DEG of the unprotected group is enough to deviate the recommendations in their favor.
No relevant pattern is reported across quartiles.
Hence, also with regard to the unbalanced groups' representation, the nodes themselves could be related to the unfairness, given that \texttt{GNNUERS} explains it by consistently pruning more their edges compared to the protected group ones.
For both FENG and LFM-1K, \texttt{GNNUERS} deletes more edges connected to high-DEG (Q4) and high-IGD (Q4) unprotected users (LFM-1K: Younger, FENG: Older), which represent those unprotected users with the most interactions and the closest ones to other unprotected user nodes.
According to Table~\ref{tab:datasets}, Older users in FENG are less represented, but their higher average DEG and IGD reflects the observations derived from Figure~\ref{fig:graph_metric_ngcf_age}.
{\color{black} On LFM-1K, the gap in DY between the protected and unprotected user nodes gets larger as the tendency to consume more popular (Q3-Q4) artists (items) increases.
Hence, the edges deleted by \texttt{GNNUERS} highlight a possible connection between popularity bias and unfairness across age groups.
This subgroup (high DEG, DY, IGD) of Younger users may have contributed to the increased popularity of certain artists.
This phenomenon may have hindered the personalization power of the system for Younger users in lower quartiles, leading to recommendations with lower overall utility.}

\begin{figure*}[!b]
    \includegraphics[width=\linewidth]{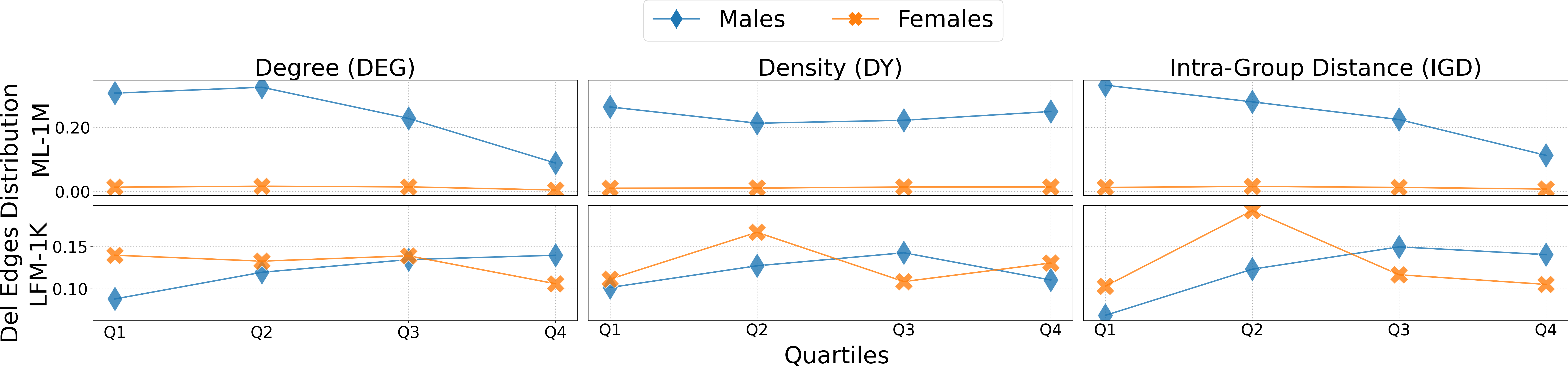}
    \caption{Deleted edges distribution (\emph{Del Edges Distribution}) over the quartiles (Q1-Q2-Q3-Q4) defined for each \emph{gender} group through sorting the nodes by each graph property. The edges were deleted applying \texttt{GNNUERS} on NGCF.}
    \Description{}
    \label{fig:graph_metric_ngcf_gender}
\end{figure*}

\noindent \textbf{Gender.} Figure~\ref{fig:graph_metric_ngcf_gender} reports the distribution of the deleted edges over the graph properties quartiles, defined for each gender group.
FENG does not include gender labels, so it could not be included in this figure.
On ML-1M, the significant difference of deleted edges between Males and Females underlines the unfairness may be affected by an over-representation of the unprotected group (Males) in terms of nodes and edges, i.e. over-DEG (according to Table~\ref{tab:datasets}).
Instead, the trend observed in the distribution of deleted edges over DEG and IGD suggests that unfairness may be influenced by the interactions of low-DEG and -IGD (Q1) Males.
Hence, we conjecture that \texttt{GNNUERS} identifies the interactions of isolated (low IGD) Males who watched fewer movies (low DEG) as a factor that may have led NGCF to favor this gender group.
Moreover, the deleted edges distribution over Males is slightly higher towards the lowest (DY Q1) and highest (DY Q4) popularity levels, restricting the \texttt{GNNUERS} edges selection to isolated Males connected with just a few of mostly niche or mainstream movies.
On LFM-1K, \texttt{GNNUERS} perturbes the interactions applying a dual effect.
Except for DY, lower values (Q1) of the graph properties report a higher amount of deleted edges connected to the unprotected (Females) nodes, while the nodes with higher values (Q4) that lost more edges are the protected (Males) ones.
Unfairness explanation is then accomplished by the simultaneous perturbation of low-DEG, low-IGD Females and high-DEG, high-IGD Males.
Removing the interactions of isolated (low IGD) Females listening to a few artists (low DEG) may have drastically affected the fairness levels, causing \texttt{GNNUERS} to inevitably perturb edges connected to nodes with the opposite properties (high-IGD, high-DEG Males) to restore fairness.
The fact that Males report a higher Gini on DY compared to Females suggests the former tend more to interact with artists of diverse popularity, which emphasizes how the modification of the artists' popularity impacts the recommendations.

% RIMOVIBILE
% \texttt{GNNUERS} on LightGCN report a lower KL divergence in all the settings compared to the other models.
% Hence, even though \texttt{GNNUERS} is more dependent on user nodes SP, RB and item nodes DEG, the defined properties in our study could not cover all the aspects that affect the \texttt{GNNUERS} edges deletion process and other graph properties could be considered in future works.

\hlbox{RQ3}{
\texttt{GNNUERS} edges selection process is significantly affected by the dataset domain. Experiments uncovered an unfairness mainly related to differences between demographic groups in interest (DEG) and closeness (IGD).
}

\section{Discussion and Conclusions} \label{sec:discuss}

% What was the problem we wanted to tackle
% How did we tried to solve the problem
%In this paper, we first depicted the state of the art on research into analysis and mitigation of unfairness in recommendation systems, into techniques that manipulate the network structure to improve fairness or explain predictions of GNNs (Section \ref{sec:related}).
In this work, we introduced \texttt{GNNUERS}, a framework to explain unfairness in GNN-based recommender systems.
\texttt{GNNUERS} aims to find the set of edges generating unfairness in GNNs recommendations through counterfactuality (Section \ref{sec:method}).
We proved the effectiveness of our method on three state-of-the-art GNN-based recommender systems and evaluated it on four real-world datasets, focusing on the disparity in recommendation utility between gender and age groups (Section \ref{sec:eval}).
In this section, we discuss the \texttt{GNNUERS} explanations interpretation, limitations and future extensions.

\subsection{Interpretation of Unfairness Explanations}

The counterfactual explanations generated by \texttt{GNNUERS} aim to support system designers and service providers to improve the fairness of recommendation models outcomes.
However, differently from other model-level explanations methods in GNNs \cite{DBLP:conf/kdd/YuanTHJ20,DBLP:conf/wsdm/HuangKMRS23}, user-item interaction networks generally consist of a significantly high number of nodes and edges, e.g., the graphs used in our experiments.
{\color{black} Moreover, edge-level explanations generated at a local scope~\cite{DBLP:conf/aistats/LucicHTRS22,DBLP:conf/pkdd/KangLB21} can be easily visualized as an edge mask over a node's ego-graph, which is significantly smaller than the entire network.}
It follows that visualizing the original graphs with the application of the edges mask selected by \texttt{GNNUERS} does not convey sufficient information to interpret the aspects affecting the prior unfairness.
Analyzing the edges as they are is therefore challenging and it is crucial to account for other features and properties of the nodes connected to such edges.

In the previous section, we leveraged the graph properties introduced in Section~\ref{subsec:graph_properties} to characterize the edges identified by \texttt{GNNUERS} as a possible source of the user unfairness.
The selected properties reflect different aspects of the GNNs architecture and of the recommendation scenario: DEG represents the users' interest in the items and the available nodes for the GNNs aggregation step, DY represents the users' tendency to interact with popular items, IGD represents the users' closeness in the network.
It follows that the insights provided by the nodes categorization of the unfairness explanations by means of the selected properties could be synthesized in a more compact and user-friendly form, e.g., a textual description.
We implicitly performed such a synthesis in Section~\ref{subsec:edge_selec_process} when we stated that \emph{``\texttt{GNNUERS} identifies the interactions of isolated (low IGD) Males who watched fewer movies (low DEG) as a factor that may have led NGCF to favor this gender group''} for ML-1M, and that \emph{``unfairness explanation is then accomplished by the simultaneous perturbation of low-DEG, low-IGD Females and high-DEG, high-IGD Males''} for LFM-1K.
Despite these statements highlight relevant aspects of the prior user unfairness in the recommendations, they could be extended by accounting for other user and item features, e.g., the movie (ML-1M) and the songs (LFM-1K) genres.
However, not all datasets include additional user/item features (e.g., INS).
Hence, a more thorough and comprehensive investigation could provide more informative findings and it will be one of the main objectives in our future works.

\subsection{Limitations and Future Works}

\begin{table}[!b]
\centering
\caption{\texttt{GNNUERS} execution timings. \emph{1/$C$ Iter. Time} stands for the timings after 1 and $C$ iterations.}
\label{tab:timings}
\resizebox{.6\linewidth}{!}{
\begin{tabular}{l|r|r|r|r}
\toprule
         & \multicolumn{1}{c|}{ML-1M} & \multicolumn{1}{c|}{FENG} & \multicolumn{1}{c|}{LFM-1K} & \multicolumn{1}{c}{INS} \\
         & \multicolumn{1}{c|}{1/$C$ Iter. Time} & \multicolumn{1}{c|}{1/$C$ Iter. Time} & \multicolumn{1}{c|}{1/$C$ Iter. Time} & \multicolumn{1}{c}{1/$C$ Iter. Time} \\
\midrule
GCMC & $\sim$15s/$\sim$300s & $\sim$40s/$\sim$1,800s & $\sim$10s/$\sim$120s & $\sim$1s/$\sim$10s \\
LightGCN & $\sim$15s/$\sim$300s & $\sim$30s/$\sim$2,700s & $\sim$2s/$\sim$60s & $\sim$1s/$\sim$30s \\
NGCF & $\sim$40s/$\sim$2,400s & $\sim$30s/$\sim$600s & $\sim$2s/$\sim$120s & $\sim$1s/$\sim$15s \\
\bottomrule
\end{tabular}
}
\end{table}

% What are the findings that we found and what will they bring to the community?
\texttt{GNNUERS} is the first step towards explaining unfairness in GNN-based recommender systems through counterfactuality, that is finding which interactions of the sensitive groups led to unfairness.
\texttt{GNNUERS} reduces the unfairness in recommendations, but with the purpose of explaining the issue, not mitigating it.
Indeed, we devised our method to seek for edges to eliminate from the original graph because of our objective of finding one of the possible sources of the unfairness, represented as the users' behavior (interactions) affecting the model outcomes.
As shown through the experiments, this procedure might lead to reduced recommendation utility for the unprotected groups, due to the fact that their interactions are deleted and identified by \texttt{GNNUERS} as compromising the recommendation fairness.
We plan to extend \texttt{GNNUERS} in the future to augment the original graph, i.e. adding edges and/or nodes, as an explanation tool, but also as a procedure to mitigate the unfairness without loss in accuracy.
%The \texttt{GNNUERS} objective function reduces the recommendation utility disparity negatively affecting the unprotected group utility more than the protected one, and, under certain conditions, increasing it for the latter group.
%\texttt{GNNUERS} edges selection process is guided by the differences between user nodes in terms of DEG and IGD, related to the user interest and closeness.

% What are the limitations of our work? Using datasets with limited size,
%Our findings are based on experiments on four datasets, each one coming from a different domain and covering a wide range of scenarios, e.g. high and low sparsity, high and low number of interactions.
%Our findings are based on experiments on four real-world data, which presents sensitive attributes for the user.
%The size of the selected datasets could still not be enough to guarantee generalization of our conclusions, especially for INS, whose edges do not cover neither 1\% of the other networks.
%However, datasets sensitive information in recommender systems literature are still limited.
%\texttt{GNNUERS} works with sparse representation of adjacency matrices, working with GNNs requires high amount of memory and computational power.
%This can impede the usage of the few datasets that provide users' protected attributes and have a greater size than the ones used in our work, e.g. Last.FM 2B \cite{DBLP:journals/ipm/MelchiorreRPBLS21}, BookCrossing \cite{DBLP:conf/www/ZieglerMKL05}.

{\color{black} \texttt{GNNUERS} represents a powerful solution to generate explanations of unfairness issues on the consumer side of recommender systems.
In order to spotlight our framework as a valuable tool to perform the targeted task under different experimental protocols, we focus on the resources usage introduced in Section~\ref{subsec:resource_usage} and discuss in depth the time requirements of \texttt{GNNUERS}.
Table~\ref{tab:timings} reports the graph generation timings after 1 and $C$ iterations on the datasets and models used in our experiments.
The generation process of the perturbed graph monitors the fairness level of the demographic groups, marking an iteration as complete when the recommendation utility for each user is measured.
Consequently, a higher number of users results in slower execution times per iteration, particularly notable in datasets including a high number of users, e.g., FENG.
Results also emphasize nuances across models, influenced by the diverse characteristics of each dataset.
For instance, GCMC timings per epoch could be affected by the larger item set of FENG and LFM-1K, while the feature transformation performed in the graph convolution operation of NGCF could be slowed by the denser adjacency matrix of ML-1M.
The overall timings varied more, since $C$, i.e., the total amount of iterations, increases until the early stopping is triggered, which reflects the generation process will result in a more accurate unfairness explanation.

Given that the main component affecting the execution times is the cardinality of the user set, processing larger batches of users can significantly reduce the time needed by \texttt{GNNUERS} to reach convergence.
Moreover, the global nature of the unfairness explanations requires \texttt{GNNUERS} to be run just once for each sensitive attribute, in contrast with local explanations that are generated for each user.
In light of these observations, \texttt{GNNUERS} linearly scale as the number of users increases.
Instead, the computational time on datasets with a large amount of interactions could only be influenced by the GNN backbone used to run \texttt{GNNUERS}, not by the graph generation process itself.}

% What are the limitations of our work? only binary attributes.
% Future works will makes use of bigger data sets, more sensitive attributes, more models and different utility function for the fairness loss
\texttt{GNNUERS} was used to explain recommendation utility unfairness among demographic subgroups in a binary setting.
Having only two subgroups better highlights the edges selection performed by our framework and offers a clearer overview of the \texttt{GNNUERS} explainability power.
Even if we considered protected attributes as binary features, they are by no means binary and the age labels binarization was performed according only to the desired final groups distribution, without any discriminatory intentions against users' age.
Nonetheless, we plan to extend \texttt{GNNUERS} in settings presenting more subgroups for each sensitive attribute, and also with different subgroups cardinality.
% , but the outcomes would be more complicated and could not be enough to apprehend the underlying cause that favored one or more of the considered demographic groups.

Moreover, given the flexibility of the framework, \texttt{GNNUERS} can be also extended to other GNN-based architectures, with different aggregation methods or GNN-layers, e.g., GAT.
%Other sensitive attributes and a larger number of demographic groups will also be considered to examine the generated explanations under different scenarios.

\bibliographystyle{ACM-Reference-Format}
\bibliography{sample-base}

\end{document}